\documentclass[10pt]{iopart}

\usepackage{iopams}  
\usepackage{geometry}
\usepackage{array}
\usepackage{graphicx}
\usepackage{booktabs}
\usepackage{gensymb}
\usepackage{hyperref}
\usepackage{makecell}

\makeatletter
\expandafter\let\csname equation*\endcsname\relax
\expandafter\let\csname endequation*\endcsname\relax
\makeatother

\hypersetup{
    colorlinks=true,
    linkcolor=blue,
    filecolor=magenta,      
    urlcolor=blue,
    citecolor=blue,
    pdftitle={Overleaf Example},
    pdfpagemode=FullScreen,
    }

\usepackage{amsmath,amssymb}
\usepackage[
backend=biber,
style=phys,
maxnames=1,
]{biblatex}
\usepackage{blindtext}
\begin{document}

\title[High-power TCV scenario for conventional and alternative divertor studies]{High-power TCV scenario for conventional and alternative divertor studies}

\author{K.~Lee$^1$, C.~Theiler$^1$, M.~Carpita$^1$, M.~Zurita$^1$,  P.~Sintre$^1$, O.~Février$^1$, F.~Pastore$^{1,2}$, H.~Reimerdes$^1$, K.~Verhaegh$^3$, M.~Winkel$^{3,4}$, D.~Brida$^2$, B.Y.K.~Brown$^1$, M.J.H. Cornelissen$^{1,3,4}$, R.~Ducker$^1$, G.~Durr-Legoupil-Nicoud$^1$, D.~Hamm$^1$, R.I.~Morgan$^1$, A.~Perek$^1$, O.~Sauter$^1$, E.~Tonello$^1$, Y.~Wang$^1$, the TCV Team$^{\dag}$ and the EUROfusion Tokamak Exploitation Team$^{\ddag}$}

\address{$^1$ \'Ecole Polytechnique F\'ed\'erale de Lausanne (EPFL), Swiss Plasma Center (SPC), CH-1015 Lausanne, Switzerland}
\address{$^2$ Max Planck Institute for Plasma Physics, Garching, Germany}
\address{$^3$ Department of Applied Physics and Science Education, Science and Technology of Nuclear Fusion, Eindhoven University of Technology, Eindhoven, Netherlands}
\address{$^4$ DIFFER - Dutch Institute for Fundamental Energy Research, Eindhoven, Netherlands}
\address{$\dag$ See author list of C. Theiler, et al., Nucl. Fusion 66 (2026) 116007.}
\address{$\ddag$ See author list of N. Vianello, et al., Nucl. Fusion 66 (2026) 116010.}

\ead{kenneth.lee@epfl.ch}

\vspace{10pt}

\begin{abstract}
Alternative divertor configurations (ADCs) must be evaluated under boundary plasma conditions approaching reactor-level values to be considered a reliable, physics-based solution for tokamak power exhaust. Most ADC experiments performed to date were at relatively low exhaust power. This work presents a high-power scenario on the TCV tokamak enabling the study of a wide variety of divertor magnetic shapes under an expanded SOL and power exhaust parameter space. The scenario is characterized by high power levels of electron cyclotron resonance heating ($2.5\,\text{MW}$ fully absorbed in a $\sim1\,\text{m}^{3}$ plasma) at high plasma current (edge safety factor $q_{95}\approx 2.5$), and low upstream separatrix densities ($n_{e,\text{u}}\approx1\times10^{19}\,\text{m}^{-3}$, Greenwald fraction $f_{\text{G}}\approx 0.1$). Stationary parallel heat fluxes up to $100\,\text{MW m}^{-2}$ are measured at the divertor target, an order of magnitude above previous TCV power exhaust studies. The obtained SOL collisionality and Lengyel detachment scaling metric lie within range of values expected in future reactors (SPARC, ITER, ARC). 

\end{abstract}

%
\vspace{2pc}
\maketitle
%
\ioptwocol

\section{Introduction}
Intensive research has been dedicated into the possibility of improving tokamak power exhaust beyond the conventional ITER-like single-null (SN) divertor \cite{pitts_physics_2019} by means of innovative magnetic shaping \cite{zohm_eu_2021}. Alternative divertor configurations (ADCs) \cite{theiler_results_2017,verhaegh_divertor_2025,soukhanovskii_advanced_2013}, also known as advanced divertors, exploit physics principles of divertor magnetic shaping to facilitate access to plasma detachment \cite{leonard_plasma_2018,krasheninnikov_physics_2017} at reduced requirements of upstream separatrix density and impurity seeding \cite{kallenbach_impurity_2013}, as well as to widen the parameter window of detached operation. The main theme of ADC research consists of understanding how various geometric parameters---such as poloidal and total flux expansion, divertor leg length, and secondary X-points---impact physical processes in divertor plasmas, in order to reliably predict how these configurations will perform in a reactor.

Divertor magnetic geometry can modify detachment in two distinct ways. The first case concerns tokamaks with a short-legged divertor and flexibility of divertor shaping, which is available in high-powered devices ($>10\,\text{MW}$) like DIII-D \cite{petrie_application_2015,soukhanovskii_developing_2018}, JET \cite{lowry_divertor_1997,loarte_plasma_1998}, and the recently upgraded AUG with its ADC-capable upper divertor \cite{lunt_proposal_2017,lunt_study_2021,putterich_overview_2026}. Generally, the cold, radiative detachment front \cite{lipschultz_sensitivity_2016} is observed to move rapidly from the target toward the X-point upon detachment onset \cite{loarte_plasma_1998,potzel_new_2014}, forming an X-point radiator (XPR) \cite{bernert_x-point_2025} (or a MARFE \cite{lipschultz_marfe_1984} that can cause a disruption). The properties of the XPR regime (e.g., stability, size of the radiation zone, ELM behavior) may be influenced by the magnetic geometry \cite{ryutov_snowflake_2015} in the vicinity of the active X-point \cite{gorno_power_2023,gorno_x-point_2024,lunt_compact_2023,reimerdes_access_2024,soukhanovskii_search_2024}.

A different phenomenology in detachment evolution emerges when the divertor leg is sufficiently long \cite{harrison_detachment_2017}. Besides pioneering work on DIII-D \cite{petrie_effect_2013}, long-legged divertor configurations have been studied mainly at TCV and MAST-U. It has been shown that the detachment front evolves continuously along the divertor leg within the detachment parameter window  \cite{theiler_results_2017,verhaegh_divertor_2025,koenders_model-based_2023,reimerdes_tcv_2017} and can be actively controlled \cite{ravensbergen_real-time_2021,koenders_model-based_2023,kool_demonstration_2025}. Further widening of the detachment window has been predicted and observed through advanced shaping with poloidal flux expansion (X-divertor \cite{takase_guidance_2001,kotschenreuther_heat_2007,kotschenreuther_magnetic_2013,theiler_results_2017}), total flux expansion (Super-X divertor \cite{valanju_super_2010,harrison_benefits_2024,carpita_parallel_2024,verhaegh_divertor_2025}) and additional X-points (X-point target divertor \cite{labombard_adx_2015,lee_x_2025}). Long-legged divertors can be combined with tight physical baffles for enhanced neutral confinement in the divertor \cite{umansky_attainment_2017,sun_performance_2023,reimerdes_implementation_2026}. The extra degrees of freedom offered by long-legged divertors present possibilities for the strong core-divertor decoupling that is desirable in a reactor. 

Many fusion power plant development programs (EU-DEMO \cite{militello_preliminary_2021,reimerdes_assessment_2020,xiang_operational_2021}, ARC \cite{kuang_conceptual_2018,wigram_performance_2019,eich_power_2026}, STEP \cite{henderson_overview_2025,osawa_solps-iter_2023}) are assessing the physics and engineering basis of ADCs in their designs. But for any novel, alternative power exhaust concept to be considered a credible solution that can reliably extrapolates to a reactor, it must be demonstrated at scrape-off layer (SOL) plasma parameters approaching reactor values \cite{labombard_adx_2015}, or at least at levels comparable to conventional divertor experiments, where the handling of parallel heat fluxes of several hundred $\text{MW m}^{-2}$ have been studied and modelled in metal-walled \cite{kallenbach_partial_2015,brunner_surface_2017,kaveeva_solps-iter_2021} and deuterium-tritium scenarios \cite{giroud_coreedge_2024} in preparation for ITER. In the case of TCV and MAST-U, which are carbon-walled devices, most ADC power exhaust experiments studied to date produced peak parallel heat fluxes $\lesssim20\,\text{MW m}^{-2}$ measured at an attached outer target \cite{lee_x_2025,raj_improved_2022,harrison_benefits_2024}, whereas reactors are expected to produce one to two orders of magnitude higher unmitigated parallel heat fluxes ($\sim \text{GW m}^{-2}$). The large extrapolation gap of ADCs is a recognized challenge, since there is currently no intermediate ITER-like step between present experiments and a reactor \cite{zohm_eu_2021}. New AUG experiments, together with the future devices DTT \cite{giruzzi_divertor_2026} and SPARC \cite{kuang_divertor_2020}, will serve as important ADC testbeds at reactor-level heat exhaust, providing key data toward closing this gap. In the meantime, present ADC experiments must be pushed to their full capacity in terms of the obtainable power exhaust challenge.

This paper presents a new high-power L-mode scenario on TCV tailored for stationary power exhaust studies across an extremely wide variety of divertor magnetic geometries. The scenario is characterized by high ECRH power at maximum absorption, high plasma current and low upstream separatrix density. The divertor is subjected to parallel heat fluxes up to $100 \, \text{MW m}^{-2}$, unprecedented in TCV power exhaust research and an order of magnitude higher than previously studies, approaching values found in larger, high-power tokamaks such as AUG and DIII-D. The SOL collisionality and the Lengyel detachment scaling metric, which together quantify the power exhaust challenge, are evaluated and found to lie within the range of those expected in reactors (SPARC, ITER, ARC). The present scenario significantly elevates the power exhaust challenge attainable in TCV, expanding the SOL parameter space accessible for conventional divertor and ADC studies.

The paper is organized as follows: section \ref{section:scenario_dev} describes the scenario development for the high-power L-mode discharge; section \ref{section:scenario_overview} provides an overview of the scenario; section \ref{section:upstream} analyzes the achieved SOL and power exhaust parameters and presents a comparison with reactors; section \ref{section:conclusion} contains the conclusion and outlook.

\section{Development of the high-power ADC scenario in TCV}\label{section:scenario_dev}
TCV (Tokamak à Configuration Variable) \cite{hofmann_creation_1994,theiler_progress_2026} is a medium-sized tokamak experiment (major radius $R_{0}=0.88\,\text{m}$, minor radius $a\approx 0.25\,\text{m}$, toroidal magnetic field $B_{0}=1.44\,\text{T}$) capable of producing a wide range of plasma shapes using a set of 16 independently powered poloidal field (PF) coils, a fully covered carbon first wall, and a highly elongated vacuum vessel \cite{theiler_results_2017}. TCV has a long history of exploring variations in divertor magnetic geometry \cite{pitts_divertor_2001,piras_snowflake_2010}. In response to the emerging interest in developing new plasma exhaust solutions, deliberate studies of ADCs have taken place in TCV over the past decade \cite{reimerdes_tcv_2017,theiler_results_2017}, in conjunction with a series of upgrades including plasma heating, divertor diagnostics, and installation of variable gas baffles \cite{reimerdes_tcv_2017-1,fasoli_tcv_2020,reimerdes_initial_2021,fevrier_divertor_2021}. To date, most conventional and alternative divertor power exhaust studies have been carried out in Ohmic L-mode \cite{harrison_detachment_2017,reimerdes_tcv_2017,theiler_results_2017,fevrier_detachment_2021,gorno_power_2023,carpita_parallel_2024,gorno_x-point_2024} or NBI-heated H-mode \cite{harrison_progress_2019,raj_improved_2022,bernert_x-point_2023,reimerdes_access_2024} scenarios. For a given plasma scenario, the difficulty of achieving detachment can be estimated by the Lengyel metric $L_{\text{m}}$ (see section \ref{subsection:lengyel} for a detailed discussion) which takes the form
\begin{equation}
    L_{\text{m}} \equiv P_{\text{SOL}}B_{0}/(R_{0} n_{e,\text{u}}^{2}).
\end{equation}
In these previous TCV exhaust experiments, the two key variables that govern divertor detachment: power entering the SOL, $P_{\text{SOL}}$, and upstream separatrix density, $n_{e,\text{u}}$, lie in the range of $P_{\text{SOL}}\lesssim1\,\text{MW}$ and $n_{e,\text{u}}\gtrsim1.5 \times 10^{19} \,\text{m}^{-3}$. Detachment is readily obtained by raising the upstream separatrix density via deuterium puffing or radiative cooling with impurity seeding \cite{pitts_experimental_1999,fevrier_nitrogen-seeded_2020}.

New high-power experiments presented in this paper significantly elevate SOL and power exhaust parameters, characterized by both absolute values (higher $P_{\text{SOL}}$ and lower $n_{e,\text{u}}$) and dimensionless or scaling metrics (lower SOL collisionality and higher Lengyel metric). Emphasis is placed on maximizing the SOL and power exhaust challenge rather than optimizing core confinement, in order to study stationary power exhaust across different divertor geometries. The main parameters that characterize the experiments are: toroidal magnetic field $B_{0}=1.4\,\text{T}$, plasma current $I_{\text{P}}=300\,\text{kA}$, edge safety factor $q_{95}\approx2.5$, total ECRH heating power $P_{\text{ECRH}}=2.5\,\text{MW}$, core line-averaged density $\langle n_{e} \rangle = 1.3\times 10^{19}\,\text{m}^{-3}$, power entering the SOL $P_{\text{SOL}}\approx 2.5\,\text{MW}$ and upstream separatrix density $n_{e,\text{u}} \approx 0.8 \times 10^{19} \,\text{m}^{-3}$. These parameters are obtained in L-mode confinement, in both reversed (ion $\nabla B$ drift pointing away from the active X-point) and forward field (ion $\nabla B$ drift pointing towards the active X-point). The analysis presented in this paper focuses on the reversed field case.

The overarching strategy of scenario development consists of accessing low plasma density, high plasma current while avoiding disruptions, and maximizing the exhaust power. Many features relevant to divertor power exhaust studies are optimized, including shaping flexibility, first-wall and baffle clearance, narrow heat flux decay width, density control, diagnostic access, maximized power absorption, and minimized core radiation.

\subsection{Variable divertor magnetic geometry}
\begin{figure*}
    \centering
    \includegraphics[width=1\linewidth]{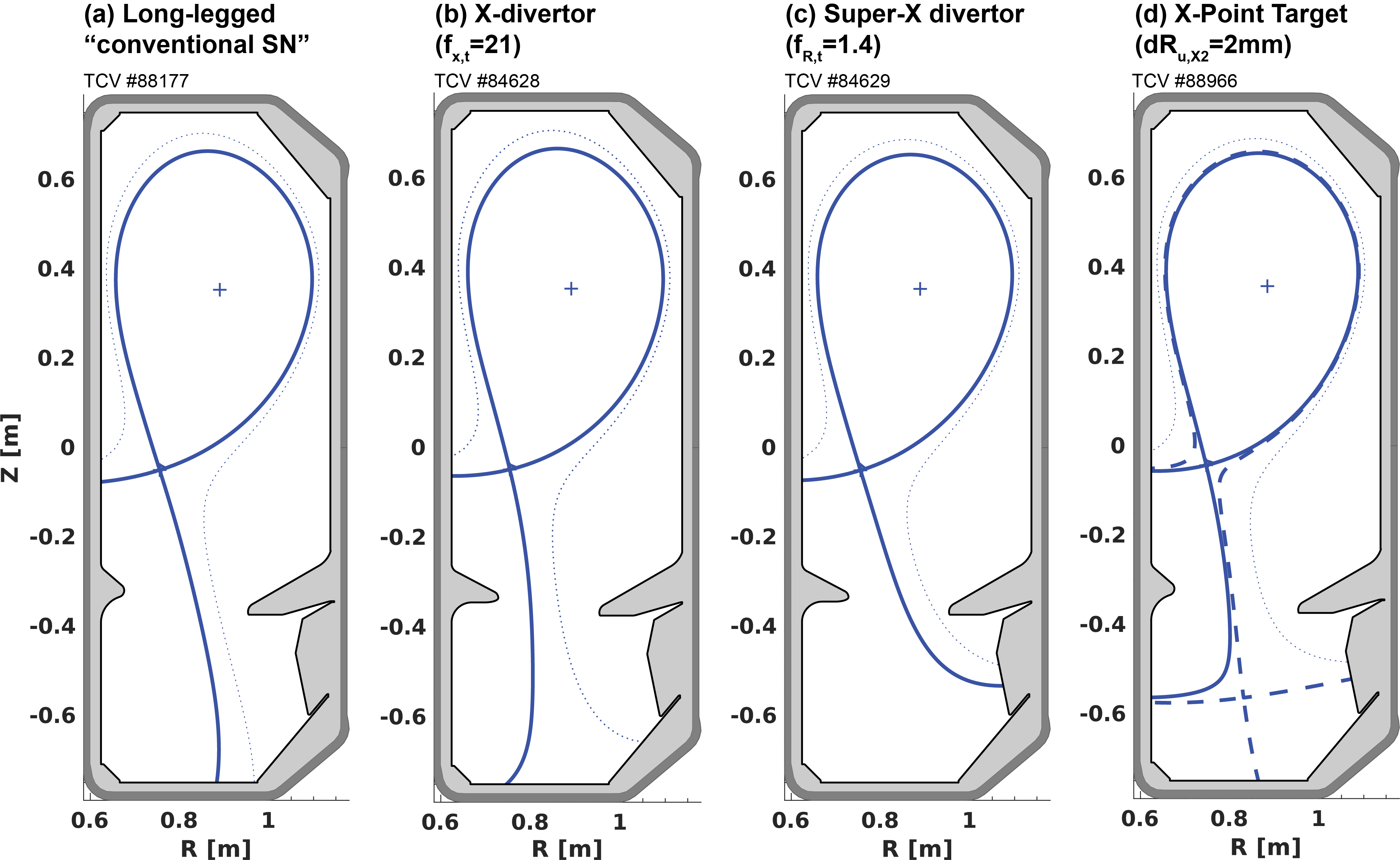}
    \caption{TCV poloidal cross-section with the LIUQE reconstructed magnetic equilibrium of (a) a long-legged conventional SN (\#88177), (b) an X-divertor (\#84628), (c) a Super-X divertor (\#84629), and (d) an X-Point Target divertor (\#88966). The divertor gas baffles are in SI-LO (Short-Inner, Long-Outer) configuration in (a, d) and LI-LO (Long-Inner, Long-Outer) configuration in (b, c).}
    \label{fig:equilibrium}
\end{figure*}

TCV can routinely execute discharges, including the present high-power scenario, across many different alternative divertor geometries \cite{theiler_results_2017}. The desired flux surface shape in TCV is configured by preparing the PF coil current traces using the inverse equilibrium solver FBT \cite{hofmann_fbt_1988,contre_kinetic_2026}. Equilibrium reconstruction is carried out using the LIUQE code \cite{moret_tokamak_2015}, which is also available in real time for magnetic feedback control of the divertor shape \cite{mele_design_2025}.

Figure~\ref{fig:equilibrium} presents examples of the achieved divertor shapes. Utilizing TCV's highly elongated ($\kappa\approx3$) vacuum vessel, the plasma can be positioned near the top to facilitate a long outer leg, with a poloidal leg length of $L_{\text{pol}}\approx0.73\,\text{m}$ approximately three times the minor radius. Long divertor legs improve the flexibility of the divertor without changing the core shape, increase proximity to the conductive shell for improved stability, and may leverage the improved detachment access observed in earlier experiments \cite{reimerdes_tcv_2017}.

Starting from a long-legged ``conventional SN", figure \ref{fig:equilibrium}(a), the divertor magnetic geometry can be flexibly configured, with the following examples (detailed definitions of geometric quantities can be found in \cite{theiler_results_2017}):
\begin{itemize}
    \item X-divertor: increased poloidal flux expansion at $f_{x,\text{t}}=\Delta R_{\text{t}}/\Delta R_{\text{u}} \approx21$ (target-to-upstream ratio of poloidal flux surface spacing), figure \ref{fig:equilibrium}(b).
    \item Super-X divertor: increased total flux expansion at $f_{R,\text{t}}\approx R_{\text{t}}/R_{\text{u}}\approx1.4$ (target-to-X-point ratio of radial position), figure \ref{fig:equilibrium}(c).
    \item X-point target divertor: secondary X-point placed at upstream-mapped X-point separation $\text{d}R_{\text{u},X2}=2\,\text{mm}$, figure \ref{fig:equilibrium}(d).
\end{itemize}

Good clearance between the separatrix and the wall (4 cm outer gap and 3.5 cm inner gap) has been obtained across all divertor configurations, minimizing main chamber interactions. The divertor configurations are also compatible with the variable gas baffles in TCV, albeit not optimized for maximum divertor neutral pressure \cite{gorno_power_2023}. Two different baffle configurations are shown in figure \ref{fig:equilibrium}.

\subsection{Real-time control of low plasma densities}\label{subsection:density}

Operating at the lowest possible density maximizes the power exhaust challenge \cite{goldston_new_2017,reinke_heat_2017} (see section \ref{section:upstream}), but also constrains plasma confinement to L-mode for two reasons: (1) H-mode plasmas typically exhibit higher core and separatrix density, which is undesirable for our purposes; and (2) in the ``low density branch", the L-H power threshold \cite{labit_lh_2025} exceeds the available heating power of TCV and is therefore inaccessible in both field directions. That said, L-mode, and thus the absence of ELMs, offer distinct advantages \cite{verhaegh_investigations_2024}: good diagnosability, density controllability, divertor magnetic geometry stability, all of which are beneficial for studying stationary power exhaust. However, the heat flux decay width $\lambda_{q,\text{u}}$ generally decreases by approximately a factor of 2 from L- to H-mode \cite{maurizio_h-mode_2021}. To compensate for the broader $\lambda_{q,\text{u}}$ in L-mode, the plasma current is pushed to high values in order to narrow $\lambda_{q,\text{u}}$ as much as possible \cite{eich_scaling_2013,maurizio_divertor_2018}. A plasma current of $300\,\text{kA}$ has been achieved, resulting in a relatively low edge safety factor of $q_{95}\approx2.5$ close to MHD stability limits.

To address the stringent requirements for precise density control, a novel density profile observer, RAPDENS-EKF \cite{pastore_model-based_2023,pastore_applications_2026}, has been employed which reconstructs the core density profile in real time from far-infrared interferometer (FIR) line integrals, Thomson scattering (TS) \cite{blanchard_thomson_2019}, and real-time LIUQE equilibrium reconstruction \cite{moret_tokamak_2015}. This is achieved using a physics-based particle transport model (RAPDENS) \cite{blanken_control-oriented_2018} combined with the extended Kalman filter (EKF) technique \cite{pastore_model-based_2023}. The equilibrium-mapped density profile enables the control target to be defined as the core line-averaged density $\langle n_{e} \rangle$ strictly within the last closed flux surface. This approach has proven robust against the systematic overestimation of $\langle n_{e}\rangle$ under different divertor shapes that afflicts standard FIR-based density control, where the FIR chord intercepting a large portion of the extended divertor leg can overestimate $\langle n_{e} \rangle$ by up to $150\%$.

Core line-averaged density of $\langle n_{e} \rangle \approx 1.3\times10^{19}\,\text{m}^{-3}$ is obtained, and corresponds to a Greenwald fraction \cite{greenwald_density_2002} of $f_{\text{G}}\approx 0.06$ and an upstream density of $n_{e,\text{u}} \approx 0.8 \times 10^{19}\,\text{m}^{-3}$. The ratio of $n_{e,\text{u}}/\langle n_{e} \rangle \sim 0.6$ is somewhat higher than the value found in Ohmic L-mode discharges ($\sim0.3$) \cite{fevrier_nitrogen-seeded_2020}. Attempts to further reduce the density resulted in the onset of a significant runaway electron population and the appearance of synchrotron radiation.

\subsection{Maximization of heating and exhaust power}\label{subsection:psol}

\begin{figure*}
    \centering
    \includegraphics[width=0.8\linewidth]{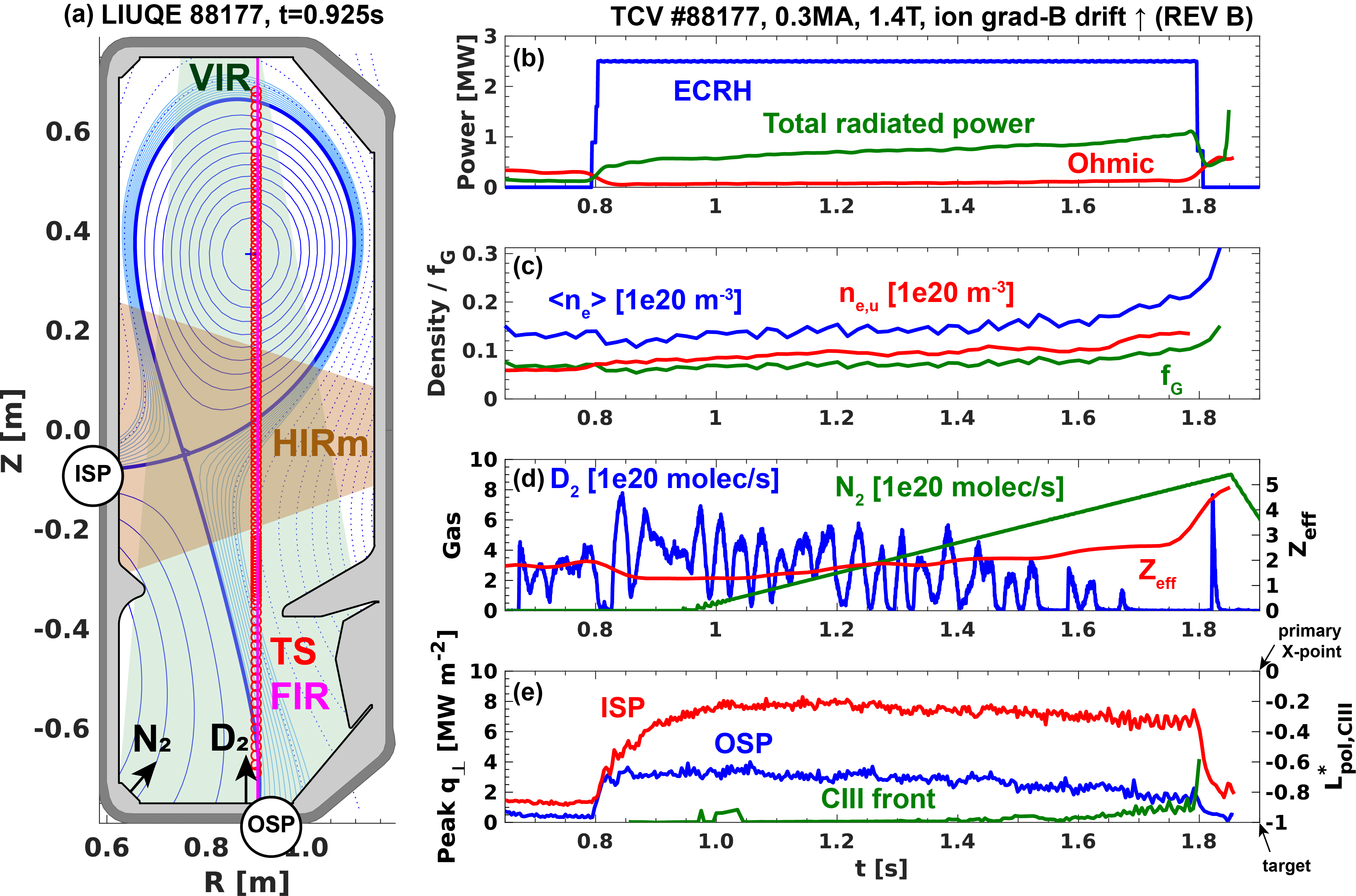}
    \caption{Overview of a high-power, reversed-field, SN discharge (\#88177). (a) LIUQE reconstruction of the magnetic equilibrium at $t=0.925\,\text{s}$. Highlighted are the locations of the inner (ISP) and outer (OSP) strike points, the vertical (VIR) and horizontal-midplane (HIRm) infrared camera fields of view, the far-infrared interferometer (FIR) feedback chord, Thomson scattering (TS) view chords, and the locations of the gas valves used to inject deuterium and nitrogen. Time traces of (b) ECRH, Ohmic and total radiated power, (c) core line-averaged density, upstream density, and Greenwald fraction; (d) deuterium and nitrogen puff rates and estimated core effective charge; (e) peak perpendicular heat flux at the inner and outer strike points, and CIII front position normalized to the leg length.}
    \label{fig:scenario}
\end{figure*}

The power entering the SOL, $P_{\text{SOL}}$, is maximized by injecting all available ECRH power at maximum core absorption while minimizing radiative losses in the core, under the condition $\langle n_{e} \rangle \approx 1.3\times 10^{19}\,\text{m}^{-3}$. This is summarized by the quasi-steady state power balance,
\begin{equation}\label{eq:power_balance}
    P_{\text{SOL}} \approx P_{\text{ECRH,abs}} + P_{\text{Ohmic}} - P_{\text{rad,core}}.
\end{equation}
Here, $P_{\text{ECRH,abs}}$ is the absorbed ECRH power. $P_{\text{Ohmic}}$ is the Ohmic heating power, and is typically small at the obtained plasma temperatures. $P_{\text{rad,core}}$ is the core radiated power.

TCV is equipped with a versatile ECRH system currently comprising two dual-frequency gyrotrons \cite{hogge_megawatt_2020} operating in second or third harmonic X-mode (X2 and X3) and one X2 gyrotron, delivering up to $2.5\,\text{MW}$ of auxiliary heating power. A main advantage of ECRH is its accessibility to plasmas at varying divertor leg length by adjusting the launcher poloidal angle. While the scenario is also available at $Z_{\text{ax}}=0\,\text{m}$, allowing NBI access, this introduces significant additional fuelling that raises the plasma density, making it incompatible with the present goal. ECRH-X2 heating is preferred over X3 because: (1) X2 is available on all three operational gyrotrons, providing the highest achievable ECRH power levels; (2) X2 offers higher heating efficiency than X3 \cite{arnoux_third_2005}; (3) X2 is compatible with low-density discharges well below the cutoff density ($n_{e,\text{cutoff}}^{\text{(X2)}}= 4.2\times10^{19}\,\text{m}^{-3}$).

Stationary L-mode confinement is maintained in both field directions under maximum ECRH-X2 power injection at $2.5\,\text{MW}$, deposited centrally with $\sim100\%$ absorbed fraction as computed by the linear ray-tracing code TORAY-GA \cite{kritz_ray_1982}. Core radiative power losses remain low ($P_{\text{rad,core}}/P_{\text{rad,tot}}\approx10-20\%$) at the operating densities in this scenario. Detachment is approached exclusively through divertor radiative losses seeded with a single low-Z impurity (nitrogen); core-radiating impurities such as argon or krypton are deliberately avoided to prevent any large increase in core radiation.

\section{Scenario overview}\label{section:scenario_overview}
This section discusses observations from a reversed field, high-power L-mode discharge. Note that while the current scenario is also achievable in forward field, many properties (e.g., radiation distribution, detachment onset, edge transport) may differ due to changes in drift directions \cite{fevrier_nitrogen-seeded_2020,jaervinen_e_2018,wang_first_2020}, and a comparison is beyond the scope of this paper. Figure \ref{fig:scenario} shows time traces of a discharge in conventional SN configuration with nitrogen seeding. Deuterium fuelling from the floor is controlled \cite{vu_integrated_2021} via RAPDENS-EKF feedback starting at $t=0.6\,\text{s}$, with a set value of $\langle n_{e} \rangle = 1.3\times10^{19}\,\text{m}^{-3}$. At $t=0.8\,\text{s}$, $2.5\,\text{MW}$ of ECRH power is injected and sustained until the discharge ends at $ t=1.8\,\text{s}$.

\subsection{Divertor heat fluxes}
\begin{figure}
    \centering
    \includegraphics[width=1\linewidth]{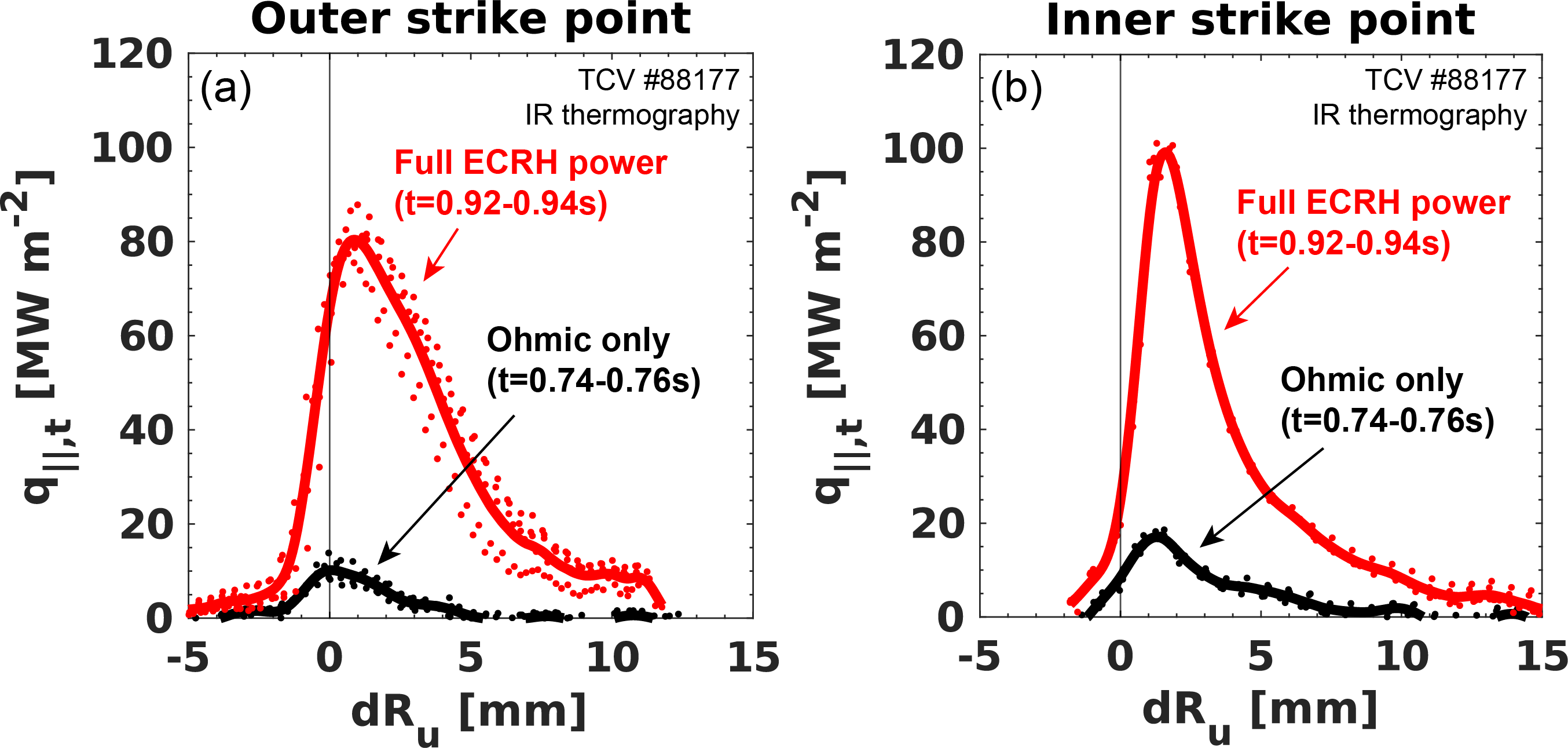}
    \caption{Target parallel heat flux profiles at the outer and inner strike points, measured by IR thermography in the conventional SN configuration (\#88177), before and after ECRH injection.}
    \label{fig:attached_heatflux}
\end{figure}

Perpendicular heat flux at the target $q_{\perp,\text{t}}$, measured by infrared thermography (IR) \cite{zurita_infrared_2026}, peaks at $3\,\text{MW m}^{-2}$ at the outer strike point (OSP) and $8\,\text{MW m}^{-2}$ at the inner strike point (ISP) during the full ECRH power phase. The target parallel heat flux $q_{\parallel,\text{t}}$ (figure \ref{fig:attached_heatflux}) is obtained from $q_{\parallel,\text{t}}=(q_{\perp,\text{t}}-q_{\text{bg}})/\sin{\alpha}$, where $q_{\text{bg}}$ is the background heat flux (assumed to be the radiation contribution) and $\alpha$ is the grazing angle of the magnetic field line on the tile surface, and peaks at $80\,\text{MW m}^{-2}$ at the OSP and $100\,\text{MW m}^{-2}$ at the ISP. The power deposited at each strike point is obtained by spatially integrating $q_{\perp,\text{t}}$ over $-5\,\text{mm}\leq\text{d}R_{\text{u}}\leq15\,\text{mm}$, yielding $620\,\text{kW}$ (OSP) and $380\,\text{kW}$ (ISP). More power is therefore deposited at the OSP, corresponding to an in-out power ratio of $\sim0.6$. Including the bolometry measurements of the total radiated power, figure \ref{fig:scenario}(b), we find that $30\text{--}40\%$ of the heating power remains unaccounted for in global power balance analysis.

Compared to the representative values of peak $q_{\parallel,\text{t}}$ at the OSP found in the previously used TCV scenarios for ADC studies: the $300\,\text{kA}$ Ohmic L-mode \cite{lee_x_2025} ($\lesssim15\,\text{MW m}^{-2}$, also shown in figure \ref{fig:attached_heatflux}), and the $170\,\text{kA}$ $1.3\,\text{MW}$ NBI-heated H-mode \cite{raj_improved_2022} ($\lesssim5\,\text{MW m}^{-2}$), the present high-power scenario elevates the divertor heat flux accessible by TCV ADCs by an order of magnitude. This also enters the parameter range of conventional divertor experiments in larger-scale facilities such as AUG and DIII-D, which typically study parallel heat fluxes of $50\text{--}150 \,\text{MW m}^{-2}$, and several hundred $\text{MW m}^{-2}$ when pushed to high heating powers \cite{eich_scaling_2013}.

In these ECRH-heated discharges, Langmuir probe \cite{fevrier_analysis_2018} measurements are found to be irregular, exhibiting multiple issues such as an order-of-magnitude underestimation of target heat flux and the absence of ion saturation under strongly negative voltage bias. Interpretation of these measurements is challenging and remains under investigation.

\subsection{Nitrogen seeding}
Nitrogen is injected from a gas valve located in the private flux region, shown in figure \ref{fig:scenario}(a), following a feedforward trace of linear ramp starting at $t=0.98\,\text{s}$ and reaches a puff rate of $9\times 10^{20}\,\text{molec. s}^{-1}$ at $t=1.85\,\text{s}$. The total amount of injected nitrogen (figure \ref{fig:upstream}(a)) is a factor of $2\text{--}3$ above typical TCV nitrogen seeding experiments \cite{fevrier_nitrogen-seeded_2020,gorno_power_2023,raj_improved_2022}. In this scenario, bolometry measurements \cite{sheikh_radcamradiation_2022} indicate that the radiation is dominated by the outer divertor, accounting for up to $65\%$ of the total radiated power. However, the total radiated power fraction  ($f_{\text{rad}}=P_{\text{rad,tot}}/P_{\text{heat}}$, where $P_{\text{heat}}=P_{\text{ECRH,abs}}+P_{\text{Ohmic}}$) is rather low, increasing from $20\%$ to $40\%$ ($P_{\text{rad,tot}}\approx0.5\text{--}1\,\text{MW}$) with nitrogen seeding.

The increase in radiated power with nitrogen seeding occurs mainly in the divertor. The core radiated power also increases, but this is offset by a simultaneous rise of Ohmic heating power estimated by LIUQE, resulting in a steady $P_{\text{SOL}}\approx2.5\,\text{MW}$ throughout the discharge (see section \ref{section:upstream}, figure \ref{fig:upstream}(a)). Initially, only about $10\text{--}15\%$ of the $P_{\text{SOL}}$ is radiated away in the boundary ($P_{\text{rad},\text{SOL}}/P_{\text{SOL}}$, where $P_{\text{rad,SOL}}$ includes radiation from the upstream SOL, X-point, and divertor); with nitrogen seeding, this rises and saturates at $25\%$. These observations indicate that the ability of nitrogen impurities to radiate in the SOL is much more limited in this scenario than in Ohmic L-mode, where $\sim60\text{--}80\%$ of $P_{\text{SOL}}$ can be radiated \cite{fevrier_nitrogen-seeded_2020}. 

The CIII (465 nm) carbon impurity emission front is used as a proxy for the low-temperature region ($T_{e}\sim7\,\text{eV}$ \cite{martinelli_spectroscopic_2023}) along the divertor to monitor the access to detachment-relevant conditions \cite{theiler_results_2017,harrison_detachment_2017}, and is tracked by the multispectral imaging diagnostic MANTIS \cite{perek_mantis_2019}. Despite intense impurity seeding, the CIII front shows little movement from the OSP, figure \ref{fig:scenario}(e). The target perpendicular heat flux shows a weaker drop at the ISP ($\sim20\%$) than at the OSP ($\sim60\%$) at the highest level of seeding. The core effective charge $Z_{\text{eff}}$, estimated from neoclassical resistivity and bootstrap current as in \cite{sauter_neoclassical_1999,sauter_erratum_2002}, increases from 1.2 to 2.6 and rises sharply to 4 at the last $50\,\text{ms}$ of the discharge, indicating a high impurity content in the core. Therefore, divertor plasma cooling to detachment-relevant temperatures by means of nitrogen seeding appears to be challenging in the present SOL regime in TCV.

\subsection{Core profiles}

\begin{figure}
    \centering
    \includegraphics[width=1\linewidth]{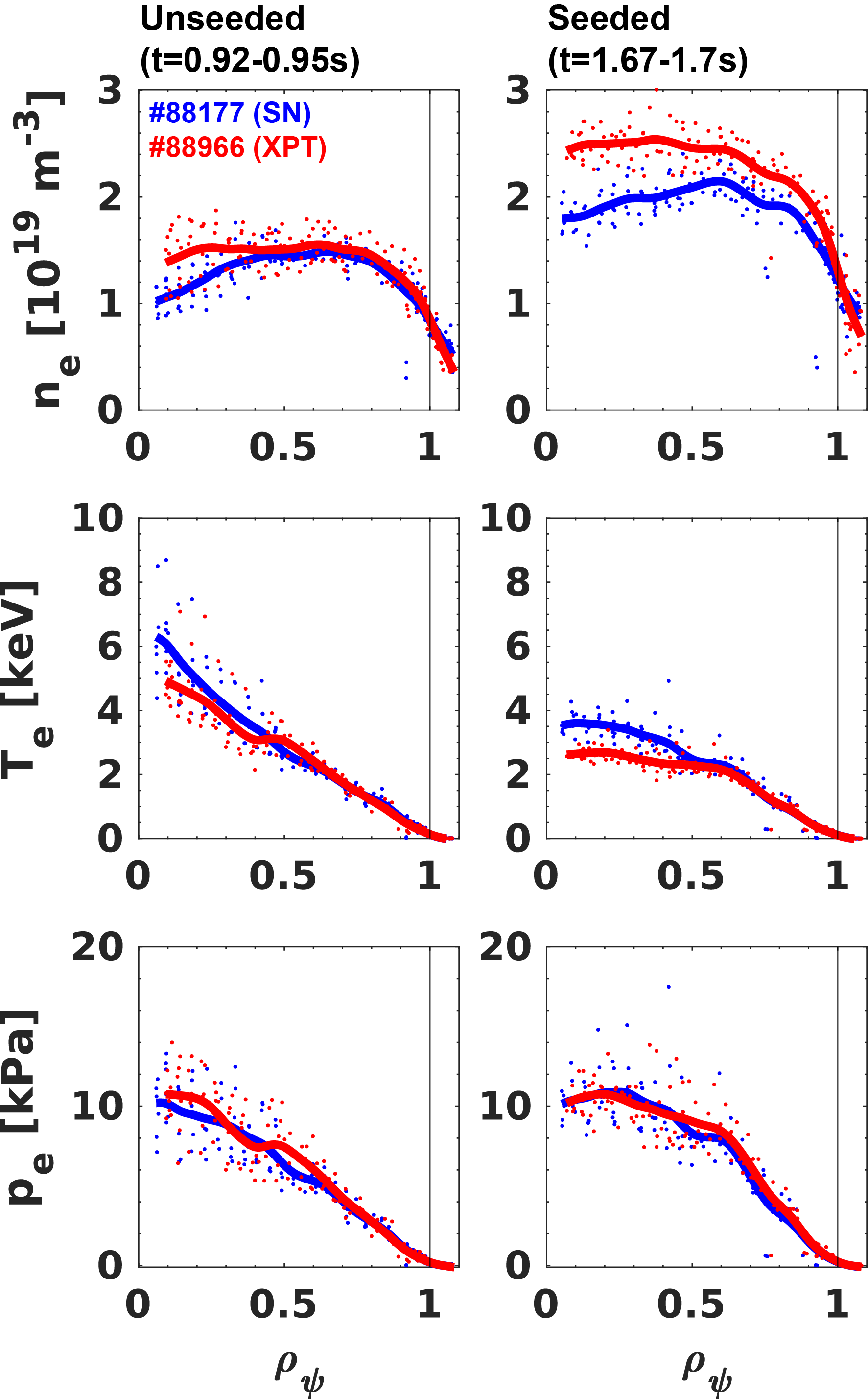}
    \caption{Core radial profiles of electron density, temperature and pressure of a conventional SN (\#88177) and an XPT (\#88966) discharge, measured by Thomson scattering during the unseeded ($t=0.92\text{--}0.95\,\text{s}$) and seeded ($t=1.67\text{--}1.7\,\text{s}$) phases, each spanning a $\sim30\,\text{ms}$ window. The vertical line indicates the position of the separatrix ($\rho_{\psi}=1$).}
    \label{fig:core_profile}
\end{figure}

Figure \ref{fig:core_profile} shows the core TS profiles of this high-power scenario in conventional SN and X-Point Target (XPT) configuration before and after seeding. The radial coordinate in these profiles is the normalized poloidal magnetic flux $\rho_{\psi}=\sqrt{(\psi-\psi_{0})/(\psi_{1}-\psi_{0})}$, where $\psi$ is the poloidal magnetic flux with $\psi_{0}$ and $\psi_{1}$ the flux at the magnetic axis and at the active X-point.

Variations in divertor magnetic geometry do not lead to a significant impact on the core profiles in TCV \cite{raj_improved_2022}. Under high central ECRH heating, the electron temperature profile is centrally peaked and reaches up to $5-6\,\text{keV}$ in the unseeded phase. The hollowing of the electron density profile in the center, known as ``density pumpout" \cite{weisen_particle_2001}, also appears. An edge pedestal is absent, as expected in L-mode.

During nitrogen seeding, the density profile is globally elevated, leading to an increase in $n_{e,\text{u}}$, and the electron temperature appears to flatten at the center ($\rho_{\psi} \leq 0.6$), reducing the peak value to around $3\,\text{keV}$. It remains inconclusive whether the density increase is caused by nitrogen fuelling, wall recycling, or changes in confinement due to nitrogen.

\section{SOL and power exhaust parameter space}\label{section:upstream}

\begin{figure*}
    \centering
    \includegraphics[width=1\linewidth]{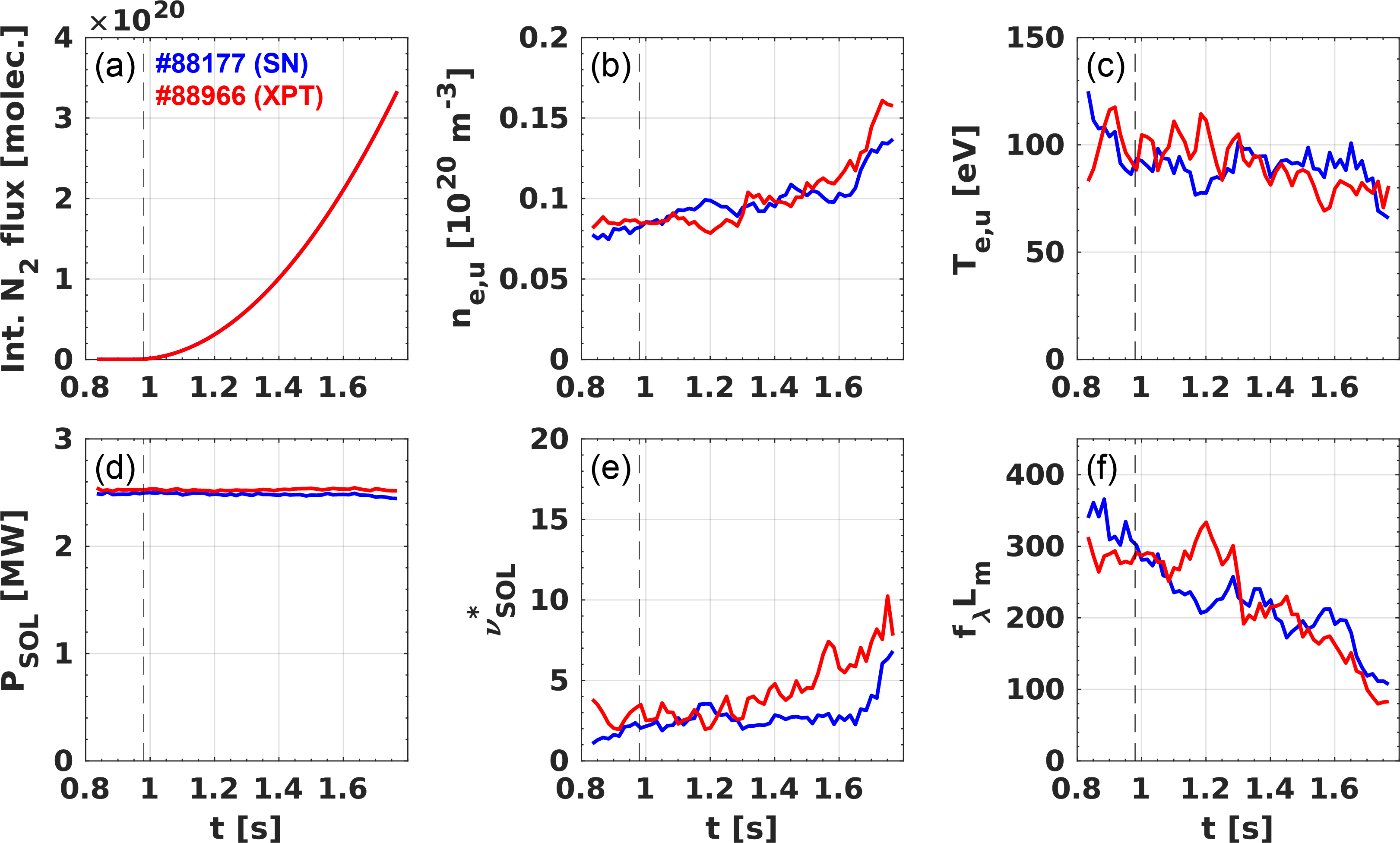}
    \caption{Time traces of upstream quantities of SN (\#88177) and XPT (\#88966) discharges: (a) time-integrated nitrogen flux; (b) upstream density; (c) upstream temperature; (d) power entering the SOL; (e) SOL collisionality; (f) L-mode adjusted Lengyel metric.}
    \label{fig:upstream}
\end{figure*}

This section presents the analysis of key upstream plasma parameters in the TCV high-power scenario, which are used as inputs to evaluate the SOL and power exhaust parameter space, quantified via the SOL collisionality and the Lengyel detachment scaling metric (defined in sections \ref{subsection:sol_collisionality} and \ref{subsection:lengyel}), and to enable comparison with reactor values. Time traces of the relevant SOL and power exhaust parameters (total nitrogen injected, upstream density and temperature, power entering the SOL, SOL collisionality, and Lengyel metric) are shown in figure \ref{fig:upstream}, where reasonably good match is obtained experimentally across different ADCs, using a conventional SN and an XPT as examples. As mentioned in section \ref{subsection:psol}, $P_{\text{SOL}}$, computed using equation (\ref{eq:power_balance}) and shown in figure \ref{fig:upstream}(d), remains approximately $2.5\,\text{MW}$ with little change throughout the nitrogen seeding ramp (figure \ref{fig:upstream}(a)). The upstream plasma parameters, SOL collisionality, and Lengyel metric are discussed in the following, along with a comparison to reactor values.

\subsection{Upstream electron density and temperature}

The upstream separatrix quantities $n_{e,\text{u}}$ and $T_{e,\text{u}}$ are extracted from Thomson scattering (TS) measurements. The TS profiles of $n_{e}$ and $T_{e}$ at the plasma edge below the magnetic axis are mapped onto the reconstructed equilibrium in $\rho_{\psi}$ coordinates. An exponential function is fitted in the interval  $0.98\leq \rho_{\psi} \leq 1.03$, corresponding to $-8\,\text{mm} \leq \text{d}R_{\text{u}} \leq 9\,\text{mm}$, which yields the separatrix values and SOL decay widths.

The time traces of the estimated $n_{e,\text{u}}$ and $T_{e,\text{u}}$ are shown in figure \ref{fig:upstream}(b) and (c), with $n_{e,\text{u}}\approx 0.8 \times 10^{19} \,\text{m}^{-3}$ and $T_{e,\text{u}}\approx 100\,\text{eV}$ during the unseeded phase. Nitrogen seeding leads to an increase in $n_{e,\text{u}}$ to $1.5\times10^{19}\,\text{m}^{-3}$ and a drop in $T_{e,\text{u}}$ to $70 \,\text{eV}$. Note that the separatrix position estimated by equilibrium reconstruction is subject to uncertainties. This has been discussed in \cite{fevrier_nitrogen-seeded_2020}, where a comparison between TS estimates and two-point model predictions, coupled with power balance arguments, found satisfactory agreement in TCV Ohmic L-mode.

\subsection{SOL collisionality}\label{subsection:sol_collisionality}
Low $n_{e,\text{u}}$ and high $T_{e,\text{u}}$ are consequential in determining the SOL transport regime, which is governed by the SOL collisionality $\nu_{\text{SOL}}^{*}$ \cite{stangeby_plasma_2000}, a dimensionless parameter defined as the ratio of the parallel connection length $L_{\parallel}$ to the electron-electron collision mean free path $\lambda_{ee}$:
\begin{equation}\label{eq:nu_SOL}
    \nu_{\text{SOL}}^{*} = \frac{L_{\parallel}}{\lambda_{ee}} \approx 10^{-16} \frac{n_{e,\text{u}}}{T_{e,\text{u}}^{2}}L_{\parallel} .
\end{equation}
Significant parallel temperature gradients, favorable for power exhaust, are expected to form at high SOL collisionality,   $\nu_{\text{SOL}}^{*}\gtrsim15$, whilst the gradients should be small when $\nu_{\text{SOL}}^{*} \lesssim 10$ \cite{stangeby_plasma_2000}. Figure \ref{fig:upstream}(e) shows the computed $\nu_{\text{SOL}}^{*}$ using the estimated $n_{e,\text{u}}$ and $T_{e,\text{u}}$, and values of $L_{\parallel}$ taken from the outboard midplane to the outer target, averaged across  $0\,\text{mm}\leq\text{d}R_{\text{u}}<2\,\text{mm}$ giving $L_{\parallel}\approx21.6\,\text{m}$ (SN) and $32.3 \,\text{m}$ (XPT). In the present high-power scenario, $\nu_{\text{SOL}}^{*} $ is generally in the range of $2\text{--}4$, indicating very low collisionality, and increases as $n_{e,\text{u}}$ increases and $T_{e,\text{u}}$ drops simultaneously during nitrogen seeding, but $\nu_{\text{SOL}}^{*}$ remains below $10$ until the end of the discharge.

To further infer the SOL transport and collisionality regime, we investigate the relationship between the heat flux decay width $\lambda_{q,\text{u}}$, and the temperature and density decay widths $\lambda_{T_{e},\text{u}}$ and $\lambda_{n_{e},\text{u}}$. In a collisional SOL, where parallel heat transport is dominated by Spitzer-H\"{a}rm (`SH') conduction, the heat flux decay width satisfies the following relation: \cite{stangeby_plasma_2000}
\begin{equation}\label{eq:lambdaq_spitzer}
    \lambda_{q,\text{u}}^{\text{SH}} = \frac{2}{7}\lambda_{T_{e},\text{u}}.
\end{equation}
In a weakly collisional SOL, instead, on the basis of flux-limited (`fl') parallel heat conduction, the heat flux decay width satisfies the following relation \cite{maurizio_h-mode_2021,stangeby_relation_2010}:
\begin{equation} \label{eq:lambdaq_fl}
    \lambda_{q,\text{u}}^{\text{fl}} = \bigg( \frac{3/2}{\lambda_{T_{e},\text{u}}} + \frac{1}{\lambda_{n_{e},\text{u}}}\bigg)^{-1}.
\end{equation}

\begin{figure}
    \centering
    \includegraphics[width=0.7\linewidth]{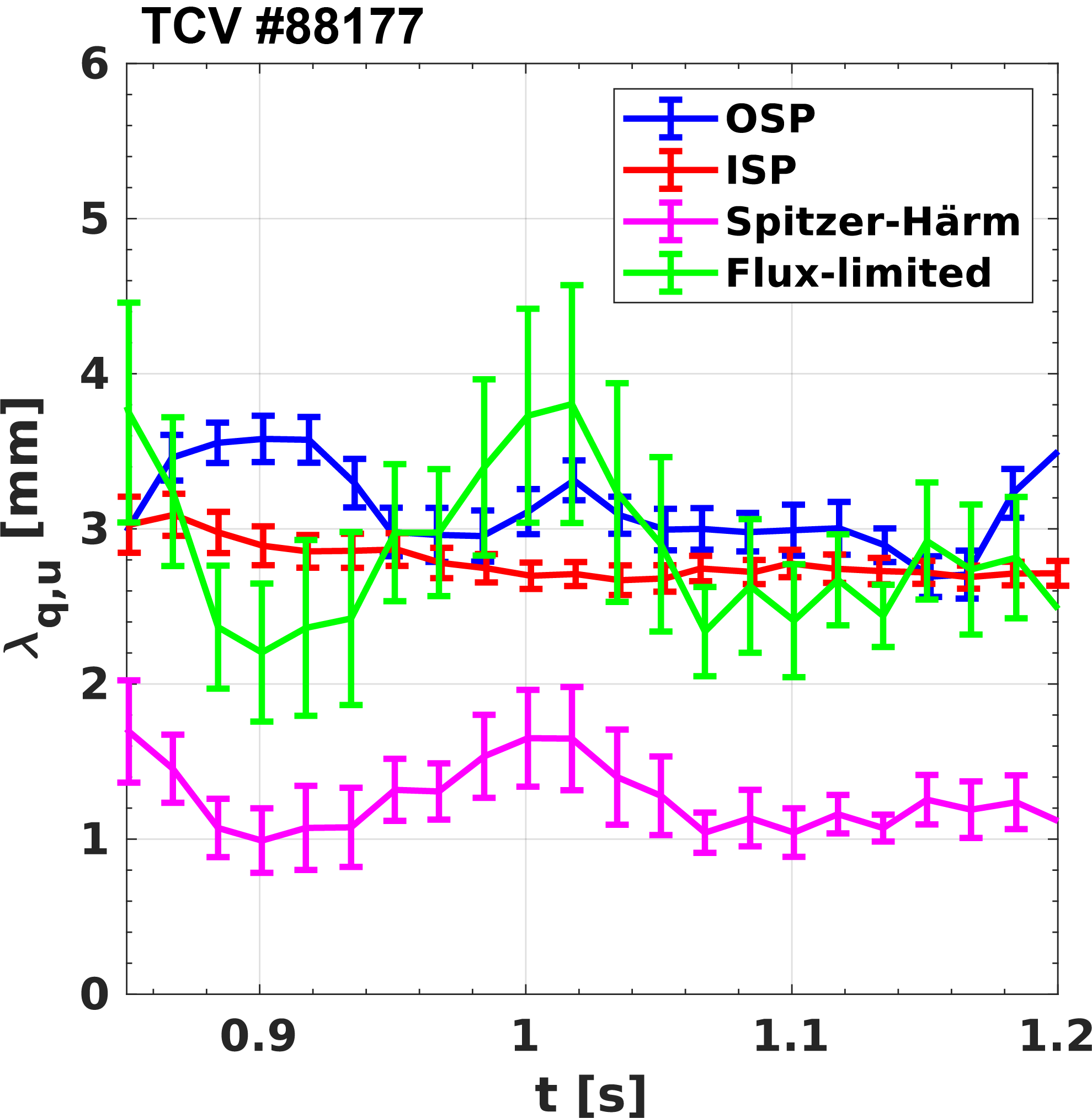}
    \caption{Time traces of heat flux decay width $\lambda_{q,\text{u}}$ in discharge \#88177, inferred from IR thermography at the outer (OSP) and inner strike point (ISP), and from the Spitzer-H\"{a}rm (equation (\ref{eq:lambdaq_spitzer})) and flux-limited (equation (\ref{eq:lambdaq_fl})) formulas.}
    \label{fig:lambdaq}
\end{figure}

Figure \ref{fig:lambdaq} compares the value $\lambda_{q,\text{u}}$ inferred from Eich fits \cite{eich_inter-elm_2011} of IR heat flux profile measurements at the OSP and ISP in SN, with estimates from the Spitzer-H\"{a}rm and flux-limited expressions (equation (\ref{eq:lambdaq_spitzer}) and (\ref{eq:lambdaq_fl})). IR measurements at both OSP and ISP suggest $\lambda_{q,\text{u}}\approx3\,\text{mm}$, similar to the reported values in TCV Ohmic L-mode  \cite{maurizio_divertor_2018}, with the OSP having a slightly higher $\lambda_{q,\text{u}}$.  The Spitzer-H\"{a}rm expression clearly underestimates $\lambda_{q,\text{u}}$ by a factor of 2. The flux-limited expression, on the other hand, yields reasonable agreement with IR, supporting the conclusion from the $\nu_{\text{SOL}}^{*}$ estimation that the SOL is weakly collisional.

\subsection{Lengyel detachment scaling metric}\label{subsection:lengyel}
The challenge in attaining divertor detachment can be quantified by the impurity concentration $c_{z}$ required to dissipate the upstream parallel heat flux. The classical analytical model by Lengyel \cite{lengyel_analysis_1981-1,body_detachment_2024} gives the scaling
\begin{equation}\label{eq:lengyel}
    c_{z} \propto q_{\parallel,\text{u}}^{8/7} n_{e,\text{u}}^{-2} L_{\parallel}^{-6/7},
\end{equation}
which highlights a particularly strong role for the upstream separatrix density $n_{e,\text{u}}$. Equation (\ref{eq:lengyel}) represents the simplest form of detachment scaling, derived from a one-dimensional power balance in which the radiative power sink removes 100\% of the electron heat conduction \footnote{Note that the derivation involves a scaling approximation of the Lengyel integral $\int_{\text{t}}^{\text{u}}\text{d}T_{e} \, L_{z} \sqrt{T_{e}} \propto T_{e,\text{u}}$ \cite{goldston_new_2017,reinke_heat_2017}, where $L_{z}$  is the cooling rate coefficient of impurity radiation.}. Further development at various levels of sophistication have been pursued in \cite{goldston_new_2017,reinke_heat_2017,body_simple_2025} and can also project a stronger scaling with $n_{e,\text{u}}$. The upstream parallel heat flux $q_{\parallel,\text{u}}$ can be expressed in terms of engineering parameters as \cite{labombard_adx_2015}
\begin{equation}\label{eq:qpar}
    q_{\parallel,\text{u}} = \frac{P_{\text{SOL}}f_{\text{odiv}}}{2\pi R_{\text{u}}\lambda_{q,\text{u}}} \frac{B_{\text{u}}}{B_{\theta,\text{u}}} \propto \frac{P_{\text{SOL}}B_{0}}{R_{0}},
\end{equation}
where $f_{\text{odiv}}$ is the fraction of $P_{\text{SOL}}$ directed to the outer divertor. The scaling $q_{\parallel,\text{u}}\sim P_{\text{SOL}}B_{0}/R_{0}$ approximates the upstream values of radial position and magnetic field with on-axis values. The empirical scaling of $\lambda_{q,\text{u}}\propto B_{\theta,\text{u}}^{-1}$ for attached inter-ELM H-mode \cite{eich_scaling_2013} is also invoked to reach the final expression. Substituting equation (\ref{eq:qpar}) into (\ref{eq:lengyel}), dropping the extra power of $1/7$ on $q_{\parallel,\text{u}}$ and excluding the connection length dependence, yields
\begin{equation}\label{eq:lm}
    c_{z} \propto L_{\text{m}} \equiv (P_{\text{SOL}}B_{0}/R_{0})/n_{e,\text{u}}^{2},
\end{equation}
where $L_{\text{m}}$ is called the Lengyel metric \cite{body_detachment_2024,wigram_exploring_2024,eich_power_2026}, a simple scaling parameter that measures the difficulty of achieving detachment as a function of machine ($B_{0}$, $R_{0}$) and scenario upstream parameters ($P_{\text{SOL}}$, $n_{e,\text{u}}$). The units are $P_{\text{SOL}} \, [\text{MW}]$, $B_{0} \,[\text{T}]$, $R_{0} \, [\text{m}]$, $n_{e,\text{u}} \, [10^{20}\,\text{m}^{-3}]$. Equation (\ref{eq:lm}) is consistent with the scaling by Goldston \cite{goldston_new_2017} up to a prefactor of an aspect ratio $R/a$ if the $L_{\parallel}$ dependence is retained. Despite its simplicity, Lengyel-like models have been shown to reproduce trends in experiments \cite{henderson_parameter_2021,body_simple_2025} and high-fidelity simulations \cite{moulton_comparison_2021}. However, it should be remarked that the global scalings may be challenged, as has been shown in cases where the effects of drifts and realistic device geometry are prominent \cite{moulton_non-monotonic_2026}. To account for the factor of 2 difference in $\lambda_{q,\text{u}}$ between L- and H-mode (also discussed in section \ref{subsection:density}), the $q_{\parallel,\text{u}}$ and $L_{\text{m}}$ scaling is adjusted with a prefactor $f_{\lambda}$, defined as $f_{\lambda}=1$ in H-mode and 0.5 in L-mode. The computed Lengyel metric in the TCV high-power scenario, shown in \ref{fig:upstream}(f), reaches values of up to 350, stays at $\sim200$ during nitrogen seeding, and eventually drops to 100 before the discharge terminates.

\subsection{Reactor relevance}\label{subsection:reactor}

\begin{table*}
    \centering
    \begin{tabular}{lccccc}\toprule
         &  \makecell{TCV\\Ohmic L-mode}&\makecell{TCV\\ECRH L-mode} &  SPARC &ITER& ARC\\\midrule
         $R_{0}$ [m]&  0.88 &0.88 &  1.85 &6.2& 4.6\\
         $B_{0}$ [T]&  1.44 &1.44 &  12.2 &5.3& 11.4\\
         $P_{\text{SOL}}$ [MW]&  0.4&2.5 &  29 &100& 120\\
         $n_{e,\text{u}}$ [$10^{20}\,\text{m}^{-3}$]&  0.15&0.1&  1 &0.44& 1\\
 $T_{e,\text{u}}$ [eV]&  40&100 & 200 &275&275\\
         $L_{\parallel}$ [m]&  20&25 &  25 &70& 120\\
 Confinement mode&  L-mode&L-mode & H-mode & H-mode&H-mode\\
 $f_{\lambda}$&   0.5&0.5& 1& 1&1\\
         $\nu_{\text{SOL}}^{*}$&  19&2.5&  6.3 &4& 15.9\\
 $f_{\lambda} \cdot P_{\text{SOL}}B_{0}/R_{0}$&  0.3&2& 191& 85&297\\
 $f_{\lambda} \cdot L_{\text{m}}$&  15&205& 191 &442&297\\ 
 $f_{\lambda} \cdot (L_{\text{m}}/L_{\parallel})$&  0.7&8.2& 7.6& 6.3&2.5\\ \bottomrule
    \end{tabular}
    \caption{Representative values used to compute the SOL collisionality  $\nu_{\text{SOL}}^{*}$ (equation (\ref{eq:nu_SOL})) and Lengyel metric $L_{\text{m}}$ (equation (\ref{eq:lm})) for TCV Ohmic L-mode \cite{fevrier_nitrogen-seeded_2020}, TCV high-power scenario (ECRH L-mode), SPARC \cite{kuang_divertor_2020,lore_evaluation_2024}, ITER \cite{pitts_physics_2019,veselova_solps-iter_2021} and ARC  \cite{eich_power_2026}. The Lengyel metric is adjusted according to the confinement mode via the prefactor $f_{\lambda}$.}
    \label{tab:Lm_compare}
\end{table*}

The obtained SOL parameters in TCV high-power scenario are compared against the reference values of future reactor-class devices, SPARC \cite{kuang_divertor_2020,lore_evaluation_2024}, ITER \cite{pitts_physics_2019,veselova_solps-iter_2021} and ARC \cite{eich_power_2026}. The previously studied TCV Ohmic L-mode \cite{fevrier_nitrogen-seeded_2020} is also included. The values are summarized in table \ref{tab:Lm_compare}. Generally, a reactor-like SOL is expected to have a higher $P_{\text{SOL}}$, a narrower $\lambda_{q,\text{u}}$, and consequently higher divertor heat fluxes. Based on Spitzer-H\"{a}rm conduction ($T_{e,\text{u}} \propto q_{\parallel,\text{u}}^{2/7}$), a higher $q_{\parallel,\text{u}}$ also implies a higher $T_{e,\text{u}}$. The overall trend is a decrease in SOL collisionality $\nu_{\text{SOL}}^{*}$ and an increase in Lengyel metric $L_{\text{m}}$, both unfavorable for power exhaust.

The magnitude of divertor parallel heat flux, estimated using $ P_{\text{SOL}}B_{0}/R_{0}$, is $\sim100$ times higher in reactor devices compared to TCV. As high-field tokamaks, the on-axis toroidal field strength $B_{0}$ of SPARC and ARC is about 8 times higher than that of TCV, which may counter the exhaust challenge \cite{goldston_new_2017} by the possibility of operating at an order of magnitude higher upstream density $(n_{e,\text{u}}\sim1\times10^{20}\,\text{m}^{-3})$.

However, the achieved Lengyel metric values in TCV L-mode discharges, $\sim 200\text{--}350$ (figure \ref{fig:upstream}(f)), are found to be of similar magnitude to SPARC ($191$) and ARC ($297$). The obtained SOL collisionality $\nu_{\text{SOL}}^{*}$ is comparable or even lower than that expected in all devices considered. ITER, by contrast, yields a very high $L_{\text{m}}$ of 442, indicating that ITER upstream parameters project onto challenging conditions for detachment. While the detachment accessibility, quantified by $L_{\text{m}}$, is defined for a given set of upstream values, the connection length $L_{\parallel}$ is also relevant, since an approximately inverse dependence of the required impurity concentration $c_{z}$ on $L_{\parallel}$ is expected (equation (\ref{eq:lengyel})). While TCV and SPARC have comparable connection lengths ($L_{\parallel}=25\,\text{m}$ in the considered cases), it is foreseen to be substantially longer in ITER and ARC ($\sim3\text{--}5\times$). In particular, ARC is designed with tightly-baffled, long-legged X-point target divertors that enable a very large connection length of $\sim120\,\text{m}$ \cite{eich_power_2026}. This also explains why a more moderate SOL collisionality of $\nu_{\text{SOL}}^{*}\approx16$ is predicted for ARC. Including a simple connection length dependence in the detachment scaling, via $L_{\text{m}}/L_{\parallel}$ (with $L_{\parallel}$ in [m]), yields a value of 8.2 for TCV high-power scenario, above those of SPARC (7.6), ITER (6.3), and ARC (2.5). The present analysis indicates that the TCV high-power scenario enters a regime of very high exhaust challenge in the SOL relevant to reactors.

A similar experimental strategy to test reactor-level exhaust is planned in SPARC's ``Advanced Divertor Mission" \cite{kuang_divertor_2020,eich_power_2026}, where the performance of alternative divertor configurations \cite{wigram_performance_2019}, such as Super-X and X-Point Target, will be evaluated under ARC-relevant $\nu_{\text{SOL}}^{*}$ and $L_{\text{m}}$ (and also the turbulence parameter $\alpha_{t}$ \cite{eich_turbulence_2020} relevant to core-edge integration).

\section{Conclusion and outlook}\label{section:conclusion}
A high-power TCV scenario suitable for the study of conventional and alternative divertor configurations (ADCs) was presented. The scenario is characterized by the injection of all available ECRH heating power ($2.5\,\text{MW}$) at maximum absorption, at high plasma current ($I_{\text{P}}=300\,\text{kA}$, $q_{95}\approx2.5$), and pushing to low density ($n_{e,\text{u}}\approx 1\times10^{19}\,\text{m}^{-3}$, $f_{\text{G}}\approx 0.1$). Under these conditions, a wide range of long-legged divertor configurations, including the X-divertor (varying poloidal flux expansion), the Super-X divertor (varying total flux expansion), and the X-Point Target divertor (additional X-points), can be realized and made accessible at variable gas baffles with a large array of boundary diagnostics. The measured target parallel heat flux were shown to approach $100\,\text{MW m}^{-2}$, an order-of-magnitude increase compared to previous TCV exhaust studies, entering the range studied in conventional divertors on larger-scale tokamak facilities. The obtained SOL collisionalities and Lengyel detachment scaling metric reached reactor-relevant values, as benchmarked against SPARC, ITER and ARC.

The expanded power exhaust operational space offered by this TCV scenario provides a platform for many new studies currently being pursued, such as power exhaust and detachment characterization in ADCs, effects of field reversal, validation against SOLPS-ITER modelling, and deployment of new plasma control schemes. In the upcoming TCV divertor upgrade \cite{reimerdes_implementation_2026}, the wall structure will be modified to implement a tightly-baffled, long-legged divertor (TBLLD), alongside the commissioning of new dual-frequency gyrotrons to further increase the total available heating power. The high-power scenario presented here will serve as a stringent experimental test for the TBLLD concept.

\section*{Acknowledgment}
This work was supported in part by the Swiss National Science Foundation. This work has been carried out within the framework of the EUROfusion Consortium, partially funded by the European Union via the Euratom Research and Training Programme (Grant Agreement No 101052200 — EUROfusion).
The Swiss contribution to this work has been funded in part by the Swiss State Secretariat for Education, Research and Innovation (SERI). Views and opinions expressed are however those of the author(s) only and do not necessarily reflect those of the European Union, the European Commission or SERI.  Neither the European Union nor the European Commission nor SERI can be held responsible for them.

\DeclareFieldFormat[article]{title}{}
\printbibliography

@article{mele_design_2025,
	title = {Design and implementation of a model-based hierarchical architecture for plasma shape control in the {TCV} tokamak},
	volume = {67},
	issn = {0741-3335, 1361-6587},
	url = {https://iopscience.iop.org/article/10.1088/1361-6587/addeee},
	doi = {10.1088/1361-6587/addeee},
	number = {6},
	urldate = {2025-06-11},
	journal = {Plasma Phys. Control. Fusion},
	author = {Mele, A and Tenaglia, A and Felici, F and Galperti, C and Carnevale, D and Coda, S and Merle, A and Pironti, A and Sauter, O and Team, The Tcv and Exploitation Team, The Eurofusion Tokamak},
	month = jun,
	year = {2025},
	pages = {065035},
}

@article{hogge_megawatt_2020,
	address = {Hefei, China},
	title = {Megawatt power generation of the dual-frequency gyrotron for {TCV} at 84 and 126 {GHz}, in long pulses},
	volume = {2254},
	url = {https://pubs.aip.org/aip/acp/article-lookup/doi/10.1063/5.0014343},
	doi = {10.1063/5.0014343},
	urldate = {2025-06-06},
	journal = {AIP Conference Proceedings},
	author = {Hogge, J.-P. and Alberti, S. and Avramidis, K. A. and Bruschi, A. and Bin, W. and Cau, F. and Cismondi, F. and Dubray, J. and Fasel, D. and Gantenbein, G. and Garavaglia, S. and Genoud, J. and Goodman, T. P. and Illy, S. and Jin, J. and Legrand, F. and Marchesin, R. and Marlétaz, B. and Masur, J. and Moro, A. and Moura, C. and Pagonakis, I. Gr. and Périal, E. and Savoldi, L. and Scherer, T. and Siravo, U. and Thumm, M. and Toussaint, M. and Tran, M.-Q.},
	year = {2020},
	pages = {090006},
}

@article{verhaegh_divertor_2025,
	title = {Divertor shaping with neutral baffling as a solution to the tokamak power exhaust challenge},
	volume = {8},
	issn = {2399-3650},
	url = {https://www.nature.com/articles/s42005-025-02121-1},
	doi = {10.1038/s42005-025-02121-1},
	language = {en},
	number = {1},
	urldate = {2025-06-02},
	journal = {Commun. Phys.},
	author = {Verhaegh, Kevin and Harrison, James and Moulton, David and Lipschultz, Bruce and Lonigro, Nicola and Osborne, Nick and Ryan, Peter and Theiler, Christian and Wijkamp, Tijs and Brida, Dominik and Cowley, Cyd and Derks, Gijs and Doyle, Rhys and Federici, Fabio and Kool, Bob and Février, Olivier and Hakola, Antti and Henderson, Stuart and Reimerdes, Holger and Thornton, Andrew and Vianello, Nicola and Wischmeier, Marco and Xiang, Lingyan and {the EUROfusion Tokamak Exploitation Team} and Abate, D. and Adamek, J. and Agostini, M. and Albert, C. and Albert Devasagayam, F. C. P. and Aleiferis, S. and Alessi, E. and Alhage, J. and Allan, S. and Allcock, J. and Alonzo, M. and Anastasiou, G. and Andersson-Sunden, E. and Angioni, C. and Anquetin, Y. and Appel, L. and Apruzzese, G. M. and Ariola, M. and Arnas, C. and Artaud, J. F. and Arter, W. and Asztalos, O. and Aucone, L. and Aumeunier, M. H. and Auriemma, F. and Ayllon, J. and Aymerich, E. and Baciero, A. and Bagnato, F. and Bähner, L. and Bairaktaris, F. and Balázs, P. and Balbinot, L. and Balboa, I. and Balden, M. and Balestri, A. and Baquero Ruiz, M. and Barberis, T. and Barcellona, C. and Bardsley, O. and Baruzzo, M. and Benkadda, S. and Bensadon, T. and Bernard, E. and Bernert, M. and Betar, H. and Bianchetti Morales, R. and Bielecki, J. and Bilato, R. and Bilkova, P. and Bin, W. and Birkenmeier, G. and Bisson, R. and Blanchard, P. and Bleasdale, A. and Bobkov, V. and Boboc, A. and Bock, A. and Bogar, K. and Bohm, P. and Bolzonella, T. and Bombarda, F. and Bonanomi, N. and Boncagni, L. and Bonfiglio, D. and Bonifetto, R. and Bonotto, M. and Borodin, D. and Borodkina, I. and Bosman, T. O. S. J. and Bourdelle, C. and Bowman, C. and Brezinsek, S. and Brida, D. and Brochard, F. and Brunet, R. and Brunetti, D. and Bruno, V. and Buchholz, R. and Buermans, J. and Bufferand, H. and Buratti, P. and Burckhart, A. and Cai, J. and Calado, R. and Caloud, J. and Cancelli, S. and Cani, F. and Cannas, B. and Cappelli, M. and Carcangiu, S. and Cardinali, A. and Carli, S. and Carnevale, D. and Carole, M. and Carpita, M. and Carralero, D. and Caruggi, F. and Carvalho, I. and Casiraghi, I. and Casolari, A. and Casson, F. J. and Castaldo, C. and Cathey, A. and Causa, F. and Cavalier, J. and Cavedon, M. and Cazabonne, J. and Cecconello, M. and Ceelen, L. and Celora, A. and Cerovsky, J. and Challis, C. D. and Chandra, R. and Chankin, A. and Chapman, B. and Chen, H. and Chernyshova, M. and Chiariello, A. G. and Chmielewski, P. and Chomiczewska, A. and Cianfarani, C. and Ciraolo, G. and Citrin, J. and Clairet, F. and Coda, S. and Coelho, R. and Coenen, J. W. and Coffey, I. H. and Colandrea, C. and Colas, L. and Conroy, S. and Contre, C. and Conway, N. J. and Cordaro, L. and Corre, Y. and Costa, D. and Costea, S. and Coster, D. and Courtois, X. and Cowley, C. and Craciunescu, T. and Croci, G. and Croitoru, A. M. and Crombe, K. and Cruz Zabala, D. J. and Cseh, G. and Czarski, T. and Da Ros, A. and Dal Molin, A. and Dalla Rosa, M. and Damizia, Y. and D’Arcangelo, O. and David, P. and De Angeli, M. and De La Cal, E. and De La Luna, E. and De Tommasi, G. and Decker, J. and Dejarnac, R. and Del Sarto, D. and Derks, G. and Desgranges, C. and Devynck, P. and Di Genova, S. and Di Grazia, L. E. and Di Siena, A. and Dicorato, M. and Diez, M. and Dimitrova, M. and Dittmar, T. and Dittrich, L. and Dominguez Palacios Durán, J. J. and Donnel, P. and Douai, D. and Dowson, S. and Doyle, S. and Dreval, M. and Drews, P. and Dubus, L. and Dumont, R. and Dunai, D. and Dunne, M. and Durif, A. and Durodie, F. and Durr-Legoupil-Nicoud, G. and Duval, B. and Dux, R. and Eich, T. and Ekedahl, A. and Elmore, S. and Ericsson, G. and Eriksson, J. and Eriksson, B. and Eriksson, F. and Ertmer, S. and Escarguel, A. and Esposito, B. and Estrada, T. and Fable, E. and Faitsch, M. and Fakhrayi Mofrad, N. and Fanni, A. and Farley, T. and Farnk, M. and Fedorczak, N. and Felici, F. and Feng, X. and Ferreira, J. and Ferreira, D. and Ferron, N. and Fevrier, O. and Ficker, O. and Field, A. R. and Figueiredo, A. and Fil, N. and Fiorucci, D. and Firdaouss, M. and Fischer, R. and Fitzgerald, M. and Flebbe, M. and Fontana, M. and Fontdecaba Climent, J. and Frank, A. and Fransson, E. and Frassinetti, L. and Frigione, D. and Futatani, S. and Futtersack, R. and Gabriellini, S. and Gadariya, D. and Galassi, D. and Galazka, K. and Galdon, J. and Galeani, S. and Gallart, D. and Gallo, A. and Galperti, C. and Gambrioli, M. and Garavaglia, S. and Garcia, J. and Garcia Munoz, M. and Gardarein, J. and Garzotti, L. and Gaspar, J. and Gatto, R. and Gaudio, P. and Gelfusa, M. and Gerardin, J. and Gerasimov, S. N. and Gerru Miguelanez, R. and Gervasini, G. and Ghani, Z. and Ghezzi, F. M. and Ghillardi, G. and Giannone, L. and Gibson, S. and Gil, L. and Gillgren, A. and Giovannozzi, E. and Giroud, C. and Giruzzi, G. and Gleiter, T. and Gobbin, M. and Goloborodko, V. and González Ganzábal, A. and Goodman, T. and Gopakumar, V. and Gorini, G. and Görler, T. and Gorno, S. and Granucci, G. and Greenhouse, D. and Grenfell, G. and Griener, M. and Gromelski, W. and Groth, M. and Grover, O. and Gruca, M. and Gude, A. and Guillemaut, C. and Guirlet, R. and Gunn, J. and Gyergyek, T. and Hagg, L. and Hakola, A. and Hall, J. and Ham, C. J. and Hamed, M. and Happel, T. and Harrer, G. and Harrison, J. and Harting, D. and Hawkes, N. C. and Heinrich, P. and Henderson, S. and Hennequin, P. and Henriques, R. and Heuraux, S. and HidalgoSalaverri, J. and Hillairet, J. and Hillesheim, J. C. and Hjalmarsson, A. and Ho, A. and Hobirk, J. and Hodille, E. and Hölzl, M. and Hoppe, M. and Horacek, J. and Horsten, N. and Horvath, L. and Houry, M. and Hromasova, K. and Huang, J. and Huang, Z. and Huber, A. and Huett, E. and Huynh, P. and Iantchenko, A. and Imrisek, M. and Innocente, P. and Ionita Schrittwieser, C. and Isliker, H. and Ivanova, P. and Ivanova Stanik, I. and Jablczynska, M. and Jacobsen, A. S. and Jacquet, P. and Jansen Van Vuuren, A. and Jardin, A. and Järleblad, H. and Järvinen, A. and Jaulmes, F. and Jensen, T. and Jepu, I. and Jessica, S. and Joffrin, E. and Johnson, T. and Juven, A. and Kalis, J. and Kappatou, A. and Karhunen, J. and Karimov, R. and Karpushov, A. N. and Kasilov, S. and Kazakov, Y. and Kazantzidis, P. V. and Keeling, D. and Kernbichler, W. and Kim, H. T. and King, D. B. and Kiptily, V. G. and Kirjasuo, A. and Kirov, K. K. and Kirschner, A. and Kit, A. and Kiviniemi, T. and Kjær, F. and Klinkby, E. and Knieps, A. and Knoche, U. and Kochan, M. and Köchl, F. and Kocsis, G. and Koenders, J. T. W. and Kogan, L. and Kolesnichenko, Y. and Kominis, Y. and Komm, M. and Kong, M. and Kool, B. and Korsholm, S. B. and Kos, D. and Koubiti, M. and Kovacic, J. and Kovtun, Y. and Kowalska-Strzeciwilk, E. and Koziol, K. and Kozulia, M. and Krämer Flecken, A. and Kreter, A. and Krieger, K. and Krutkin, O. and Kudlacek, O. and Kumar, U. and Kumpulainen, H. and Kushoro, M. H. and Kwiatkowski, R. and La Matina, M. and Labit, B. and Lacquaniti, M. and Laguardia, L. and Lainer, P. and Lang, P. and Larsen, M. and Laszynska, E. and Lawson, K. D. and Lazaros, A. and Lazzaro, E. and Lee, M. Y. K. and Leerink, S. and Lennholm, M. and Lerche, E. and Liang, Y. and Lier, A. and Likonen, J. and Linder, O. and Lipschultz, B. and Listopad, A. and Litaudon, X. and LitherlandSmith, E. and Liuzza, D. and Loarer, T. and Lomas, P. J. and Lombardo, J. and Lonigro, N. and Lorenzini, R. and Lowry, C. and Luda Di Cortemiglia, T. and LudvigOsipov, A. and Lunt, T. and Lutsenko, V. and Macusova, E. and Mäenpää, R. and Maget, P. and Maggi, C. F. and Mailloux, J. and Makarov, S. and Malinowski, K. and Manas, P. and Mancini, A. and Mancini, D. and Mantica, P. and Mantsinen, M. and Manyer, J. and Maraschek, M. and Marceca, G. and Marcer, G. and Marchetto, C. and Marchioni, S. and Mariani, A. and Marin, M. and Markl, M. and Markovic, T. and Marocco, D. and Marsden, S. and Martellucci, L. and Martin, P. and Martin, C. and Martinelli, F. and Martinelli, L. and Martin-Solis, J. R. and Martone, R. and Maslov, M. and Masocco, R. and Mattei, M. and Matthews, G. F. and Matveev, D. and Matveeva, E. and Mayoral, M. L. and Mazon, D. and Mazzi, S. and Mazzotta, C. and McArdle, G. and McDermott, R. and McKay, K. and Meigs, A. G. and Meineri, C. and Mele, A. and Menkovski, V. and Menmuir, S. and Merle, A. and Meyer, H. and Mikszuta-Michalik, K. and Milanesio, D. and Militello, F. and Milocco, A. and Miron, I. G. and Mitchell, J. and Mitteau, R. and Mitterauer, V. and Mlynar, J. and Moiseenko, V. and Molna, P. and Mombelli, F. and Monti, C. and Montisci, A. and Morales, J. and Moreau, P. and Moret, J. M. and Moro, A. and Moulton, D. and Mulholland, P. and Muraglia, M. and Murari, A. and Muraro, A. and Muscente, P. and Mykytchuk, D. and Nabais, F. and Nakeva, Y. and Napoli, F. and Nardon, E. and Nave, M. F. and Nem, R. D. and Nielsen, A. and Nielsen, S. K. and Nocente, M. and Nouailletas, R. and Nowak, S. and Nyström, H. and Ochoukov, R. and Offeddu, N. and Olasz, S. and Olde, C. and Oliva, F. and Oliveira, D. and Oliver, H. J. C. and Ollus, P. and Ongena, J. and Orsitto, F. P. and Osborne, N. and Otin, R. and Oyola Dominguez, P. and Palade, D. I. and Palomba, S. and Pan, O. and Panadero, N. and Panontin, E. and Papadopoulos, A. and Papagiannis, P. and Papp, G. and Parail, V. V. and Pardanaud, C. and Parisi, J. and Parrott, A. and Paschalidis, K. and Passoni, M. and Pastore, F. and Patel, A. and Patel, B. and Pau, A. and Pautasso, G. and Pavlichenko, R. and Pawelec, E. and Pegourie, B. and Pelka, G. and Peluso, E. and Perek, A. and Perelli Cippo, E. and Perez Von Thun, C. and Petersson, P. and Petravich, G. and Peysson, Y. and Piergotti, V. and Pigatto, L. and Piron, C. and Piron, L. and Pironti, A. and Pisano, F. and Plank, U. and Ploeckl, B. and Plyusnin, V. and Podolnik, A. and Poels, Y. and Pokol, G. and Poley, J. and Por, G. and Poradzinski, M. and Porcelli, F. and Porte, L. and Possieri, C. and Poulsen, A. and Predebon, I. and Pucella, G. and Pueschel, M. and Puglia, P. and Putignano, O. and Pütterich, T. and Quadri, V. and Quercia, A. and Rabinski, M. and Radovanovic, L. and Ragona, R. and Raj, H. and Rasinski, M. and Rasmussen, J. and Ratta, G. and Ratynskaia, S. and Rayaprolu, R. and Rebai, M. and Redl, A. and Rees, D. and Refy, D. and Reich, M. and Reimerdes, H. and Reman, B. C. G. and Renders, O. and Reux, C. and Ricci, D. and Richou, M. and Rienacker, S. and Rigamonti, D. and Rigollet, F. and Rimini, F. G. and Ripamonti, D. and Rispoli, N. and Rivals, N. and Rivero Rodriguez, J. F. and Roach, C. and Rocchi, G. and Rode, S. and Rodrigues, P. and Romazanov, J. and Romero Madrid, C. F. and Rosato, J. and Rossi, R. and Rubino, G. and Rueda, J. Rueda and Ruiz, J. Ruiz and Ryan, P. and Ryan, D. and Saarelma, S. and Sabot, R. and Salewski, M. and Salmi, A. and Sanchis, L. and Sand, A. and Santos, J. and Särkimäki, K. and Sassano, M. and Sauter, O. and Schettini, G. and Schmuck, S. and Schneider, P. and Schoonheere, N. and Schramm, R. and Schrittwieser, R. and Schuster, C. and Schwarz, N. and Sciortino, F. and Scotto d’Abusco, M. and Scully, S. and Selce, A. and Senni, L. and Senstius, M. and Sergienko, G. and Sharapov, S. E. and Sharma, R. and Shaw, A. and Sheikh, U. and Sias, G. and Sieglin, B. and Silburn, S. A. and Silva, C. and Silva, A. and Silvagni, D. and Simmendefeldt Schmidt, B. and Simons, L. and Simpson, J. and Singh, L. and Sipilä, S. and Siusko, Y. and Smith, S. and Snicker, A. and Solano, E. R. and Solokha, V. and Sos, M. and Sozzi, C. and Spineanu, F. and Spizzo, G. and Spolaore, M. and Spolladore, L. and Srinivasan, C. and Stagni, A. and Stancar, Z. and Stankunas, G. and Stober, J. and Strand, P. and Stuart, C. I. and Subba, F. and Sun, G. Y. and Sun, H. J. and Suttrop, W. and Svoboda, J. and Szepesi, T. and Szepesi, G. and Tal, B. and Tala, T. and Tamain, P. and Tardini, G. and Tardocchi, M. and Taylor, D. and Telesca, G. and Tenaglia, A. and Terra, A. and Terranova, D. and Testa, D. and Theiler, C. and Tholerus, E. and Thomas, B. and Thoren, E. and Thornton, A. and Thrysoe, A. and Tichit, Q. and Tierens, W. and Titarenko, A. and Tolias, P. and Tomasina, E. and Tomes, M. and Tonello, E. and Tookey, A. and Toscano Jiménez, M. and Tsironis, C. and Tsitrone, E. and Tsui, C. and Tykhyy, A. and Ugoletti, M. and Usoltseva, M. and Valcarcel, D. F. and Valentini, A. and Valisa, M. and Vallar, M. and Valovic, M. and Valvis, S. I. and Van Berkel, M. and Van Eester, D. and Van Mulders, S. and Van Rossem, M. and Vann, R. and Vanovac, B. and Valera Rodriguez, J. and Varje, J. and Vartanian, S. and Vecsei, M. and Velarde Gallardo, L. and Veranda, M. and Verdier, T. and Verdoolaege, G. and Verhaegh, K. and Vermare, L. and Verona Rinati, G. and Vianello, N. and Vicente, J. and Viezzer, E. and Vignitchouk, L. and Villone, F. and Vincent, B. and Vincenzi, P. and Vlad, M. O. and Vogel, G. and Voitsekhovitch, I. and Voldiner, I. and Vondracek, P. and Vu, N. M. T. and Vuoriheimo, T. and Wade, C. and Wang, E. and Wauters, T. and Weiland, M. and Weisen, H. and Wendler, N. and Weston, D. and Widdowson, A. and Wiesen, S. and Wiesenberger, M. and Wijkamp, T. and Willensdorfer, M. and Wilson, T. and Wischmeier, M. and Wojenski, A. and Wuethrich, C. and Wyss, I. and Xiang, L. and Xu, S. and Yadykin, D. and Yakovenko, Y. and Yang, H. and Yanovskiy, V. and Yi, R. and Zaar, B. and Zadvitskiy, G. and Zakharov, L. and Zanca, P. and Zarzoso, D. and Zayachuk, Y. and Zebrowski, J. and Zerbini, M. and Zestanakis, P. and Zimmermann, B. and Zlobinski, M. and Zohar, A. and Zotta, V. K. and Zou, X. and Zuin, M. and Zurita, M. and Zychor, I. and {the MAST Upgrade Team} and Harrison, J. R. and Aboutaleb, A. and Ahmed, S. and Aljunid, M. and Allan, S. Y. and Anand, H. and Andrew, Y. and Appel, L. C. and Ash, A. and Ashton, J. and Bachmann, O. and Barnes, M. and Barrett, B. and Baver, D. and Beckett, D. and Bennett, J. and Berkery, J. and Boeglin, W. and Bradley, J. and Browning, P. K. and Bryant, P. and Bryant, J. and Buchanan, J. and Bulmer, N. and Carruthers, A. and Cecconello, M. and Chen, Z. P. and Clark, J. and Coy, M. and Crocker, N. and Cunningham, G. and Cziegler, I. and Da Assuncao, T. and Damizia, Y. and Davies, P. and Day, I. E. and Derks, G. L. and Dixon, S. and Doyle, R. and Dreval, M. and Duval, B. P. and Eagles, T. and Edmond, J. and El-Haroun, H. and Elmore, S. D. and Enters, Y. and Federici, F. and Fitzgerald, I. and Fitzpatrick, R. and Fuller, W. and Gahle, D. and Galdon-Quiroga, J. and Gee, S. and Gheorghiu, T. and Gibson, K. J. and Hall-Chen, V. H. and Harrison, R. and Henderson, S. S. and Hickling, C. and Hnat, B. and Howlett, L. and Hughes, J. and Hussain, R. and Imada, K. and Jepson, P. and Kandan, B. and Katramados, I. and Kazakov, Y. O. and King, D. and King, R. and Kirk, A. and Knolker, M. and Kochan, M. and Kool, B. and Kotschenreuther, M. and Lees, M. and Leonard, A. W. and Liddiard, G. and Liu, Y. Q. and Lomanowski, B. A. and Lore, J. and Lovell, J. and Mahajan, S. and Maiden, F. and Man-Friel, C. and Mansfield, F. and Martin, R. and Mazzi, S. and McAdams, R. and McClements, K. G. and McClenaghan, J. and McConville, D. and McKay, K. and McKnight, C. and McKnight, P. and McLean, A. and McMillan, B. F. and McShee, A. and Measures, J. and Mehay, N. and Michael, C. A. and Morbey, D. and Mordijck, S. and Myatra, O. and Nelson, A. O. and Nicassio, M. and O’Mullane, M. G. and Osborne, T. and Osborne, N. and Parr, E. and Parry, B. and Patel, B. S. and Payne, D. and Paz-Soldan, C. and Phelps, A. and Piron, L. and Prechel, G. and Price, M. and Pritchard, B. and Proudfoot, R. and Rhodes, T. and Richardson, P. and Riquezes, J. and Rivero-Rodriguez, J. F. and Roach, C. M. and Robson, M. and Ronald, K. and Rose, E. and Sabbagh, S. and Sarwar, R. and Saunders, P. and Scannell, R. and Schuett, T. and Seath, R. and Shi, P. and Simmonds, M. and Smith, J. and Smith, A. and Soukhanovskii, V. A. and Speirs, D. and Staebler, G. and Stephen, R. and Stevenson, P. and Stobbs, J. and Stott, M. and Stroud, C. and Tame, C. and Thomas-Davies, N. and Thornton, A. J. and Tobin, M. and Vann, R. G. L. and Velarde, L. and Vincent, C. and Voss, G. and Warr, M. and Wehner, W. and Wijkamp, T. A. and Wilkins, D. and Williams, T. and Wilson, H. R. and Wong, H. and Wood, M. and Zamkovska, V.},
	month = may,
	year = {2025},
	pages = {215},
}

@article{pastore_model-based_2023,
	title = {Model-based electron density estimation using multiple diagnostics on {TCV}},
	volume = {192},
	issn = {09203796},
	url = {https://linkinghub.elsevier.com/retrieve/pii/S0920379623001990},
	doi = {10.1016/j.fusengdes.2023.113615},
	language = {en},
	urldate = {2024-06-10},
	journal = {Fusion Eng. Des.},
	author = {Pastore, F. and Felici, F. and Bosman, T.O.S.J. and Galperti, C. and Sauter, O. and Vincent, B. and Vu, N.M.T.},
	month = jul,
	year = {2023},
	pages = {113615},
}

@article{sauter_neoclassical_1999,
	title = {Neoclassical conductivity and bootstrap current formulas for general axisymmetric equilibria and arbitrary collisionality regime},
	volume = {6},
	issn = {1070-664X, 1089-7674},
	url = {https://pubs.aip.org/pop/article/6/7/2834/464899/Neoclassical-conductivity-and-bootstrap-current},
	doi = {10.1063/1.873240},
	language = {en},
	number = {7},
	urldate = {2025-05-08},
	journal = {Physics of Plasmas},
	author = {Sauter, O. and Angioni, C. and Lin-Liu, Y. R.},
	month = jul,
	year = {1999},
	pages = {2834--2839},
}

@article{lee_x_2025,
	title = {X -{Point} {Target} {Radiator} {Regime} in {Tokamak} {Divertor} {Plasmas}},
	volume = {134},
	issn = {0031-9007, 1079-7114},
	url = {https://link.aps.org/doi/10.1103/PhysRevLett.134.185102},
	doi = {10.1103/PhysRevLett.134.185102},
	language = {en},
	number = {18},
	urldate = {2025-05-08},
	journal = {Phys. Rev. Lett.},
	author = {Lee, K. and Theiler, C. and Carpita, M. and Février, O. and Perek, A. and Zurita, M. and Brida, D. and Ducker, R. and Durr-Legoupil-Nicoud, G. and Duval, B. P. and Gorno, S. and Hamm, D. and Oliveira, D. S. and Pastore, F. and Pedrini, M. and Reimerdes, H. and Simons, L. and Tonello, E. and Verhaegh, K. and Wang, Y. and Wüthrich, C. and {the TCV Team and the EUROfusion Tokamak Exploitation Team}},
	month = may,
	year = {2025},
	pages = {185102},
}

@article{osawa_solps-iter_2023,
	title = {{SOLPS}-{ITER} analysis of a proposed {STEP} double null geometry: impact of the degree of disconnection on power-sharing},
	volume = {63},
	issn = {0029-5515, 1741-4326},
	shorttitle = {{SOLPS}-{ITER} analysis of a proposed {STEP} double null geometry},
	url = {https://iopscience.iop.org/article/10.1088/1741-4326/acd863},
	doi = {10.1088/1741-4326/acd863},
	number = {7},
	urldate = {2025-04-25},
	journal = {Nucl. Fusion},
	author = {Osawa, R.T. and Moulton, D. and Newton, S.L. and Henderson, S.S. and Lipschultz, B. and Hudoba, A.},
	month = jul,
	year = {2023},
	pages = {076032},
}

@article{body_detachment_2024,
	title = {Detachment scalings derived from {1D} scrape-off-layer simulations},
	volume = {41},
	issn = {23521791},
	url = {https://linkinghub.elsevier.com/retrieve/pii/S2352179124002424},
	doi = {10.1016/j.nme.2024.101819},
	language = {en},
	urldate = {2025-04-25},
	journal = {Nuclear Materials and Energy},
	author = {Body, Thomas and Eich, Thomas and Kuang, Adam and Looby, Tom and Kryjak, Mike and Dudson, Ben and Reinke, Matthew},
	month = dec,
	year = {2024},
	pages = {101819},
}

@article{lipschultz_marfe_1984,
	title = {Marfe: an edge plasma phenomenon},
	volume = {24},
	issn = {0029-5515, 1741-4326},
	shorttitle = {Marfe},
	url = {https://iopscience.iop.org/article/10.1088/0029-5515/24/8/002},
	doi = {10.1088/0029-5515/24/8/002},
	number = {8},
	urldate = {2024-03-27},
	journal = {Nucl. Fusion},
	author = {Lipschultz, B. and LaBombard, B. and Marmar, E.S. and Pickrell, M.M. and Terry, J.L. and Watterson, R. and Wolfe, S.M.},
	month = aug,
	year = {1984},
	pages = {977--988},
}

@article{soukhanovskii_advanced_2013,
	title = {Advanced divertor configurations with large flux expansion},
	volume = {438},
	issn = {00223115},
	url = {https://linkinghub.elsevier.com/retrieve/pii/S0022311513000238},
	doi = {10.1016/j.jnucmat.2013.01.015},
	language = {en},
	urldate = {2024-02-01},
	journal = {Journal of Nuclear Materials},
	author = {Soukhanovskii, V.A. and Bell, R.E. and Diallo, A. and Gerhardt, S. and Kaye, S. and Kolemen, E. and LeBlanc, B.P. and McLean, A. and Menard, J.E. and Paul, S.F. and Podesta, M. and Raman, R. and Ryutov, D.D. and Scotti, F. and Kaita, R. and Maingi, R. and Mueller, D.M. and Roquemore, A.L. and Reimerdes, H. and Canal, G.P. and Labit, B. and Vijvers, W. and Coda, S. and Duval, B.P. and Morgan, T. and Zielinski, J. and De Temmerman, G. and Tal, B.},
	month = jul,
	year = {2013},
	pages = {S96--S101},
}

@article{jaervinen_e_2018,
	title = {E × {B} {Flux} {Driven} {Detachment} {Bifurcation} in the {DIII}-{D} {Tokamak}},
	volume = {121},
	issn = {0031-9007, 1079-7114},
	url = {https://link.aps.org/doi/10.1103/PhysRevLett.121.075001},
	doi = {10.1103/PhysRevLett.121.075001},
	language = {en},
	number = {7},
	urldate = {2023-12-07},
	journal = {Phys. Rev. Lett.},
	author = {Jaervinen, A. E. and Allen, S. L. and Eldon, D. and Fenstermacher, M. E. and Groth, M. and Hill, D. N. and Leonard, A. W. and McLean, A. G. and Porter, G. D. and Rognlien, T. D. and Samuell, C. M. and Wang, H. Q.},
	month = aug,
	year = {2018},
	pages = {075001},
}

@article{blanchard_thomson_2019,
	title = {Thomson scattering measurements in the divertor region of the {TCV} {Tokamak} plasmas},
	volume = {14},
	issn = {1748-0221},
	url = {https://iopscience.iop.org/article/10.1088/1748-0221/14/10/C10038},
	doi = {10.1088/1748-0221/14/10/C10038},
	number = {10},
	urldate = {2023-03-27},
	journal = {J. Inst.},
	author = {Blanchard, P. and Andrebe, Y. and Arnichand, H. and Agnello, R. and Antonioni, S. and Couturier, S. and Decker, J. and D`Exaerde, T. De Kerchove and Duval, B.P. and Furno, I. and Isoz, P.-F. and Lavanchy, P. and Llobet, X. and Marlétaz, B. and Masur, J.},
	month = oct,
	year = {2019},
	pages = {C10038--C10038},
}

@book{stangeby_plasma_2000,
	edition = {0},
	title = {The {Plasma} {Boundary} of {Magnetic} {Fusion} {Devices}},
	isbn = {978-0-367-80148-9},
	url = {https://www.taylorfrancis.com/books/9781420033328},
	doi = {10.1201/9780367801489},
	language = {en},
	urldate = {2023-03-27},
	publisher = {CRC Press},
	author = {Stangeby, P.C},
	month = jan,
	year = {2000},
}

@article{fevrier_nitrogen-seeded_2020,
	title = {Nitrogen-seeded divertor detachment in {TCV} {L}-mode plasmas},
	volume = {62},
	issn = {0741-3335, 1361-6587},
	url = {https://iopscience.iop.org/article/10.1088/1361-6587/ab6b00},
	doi = {10.1088/1361-6587/ab6b00},
	number = {3},
	urldate = {2023-03-25},
	journal = {Plasma Phys. Control. Fusion},
	author = {Février, O and Theiler, C and Harrison, J R and Tsui, C K and Verhaegh, K and Wüthrich, C and Boedo, J A and De Oliveira, H and Duval, B P and Labit, B and Lipschultz, B and Maurizio, R and Reimerdes, H and {the TCV Team} and {the EUROfusion MST1 Team}},
	month = mar,
	year = {2020},
	pages = {035017},
}

@article{kallenbach_partial_2015,
	title = {Partial detachment of high power discharges in {ASDEX} {Upgrade}},
	volume = {55},
	issn = {0029-5515, 1741-4326},
	url = {https://iopscience.iop.org/article/10.1088/0029-5515/55/5/053026},
	doi = {10.1088/0029-5515/55/5/053026},
	number = {5},
	urldate = {2023-03-25},
	journal = {Nucl. Fusion},
	author = {Kallenbach, A. and Bernert, M. and Beurskens, M. and Casali, L. and Dunne, M. and Eich, T. and Giannone, L. and Herrmann, A. and Maraschek, M. and Potzel, S. and Reimold, F. and Rohde, V. and Schweinzer, J. and Viezzer, E. and Wischmeier, M. and {the ASDEX Upgrade Team}},
	month = may,
	year = {2015},
	pages = {053026},
}

@article{reimerdes_assessment_2020,
	title = {Assessment of alternative divertor configurations as an exhaust solution for {DEMO}},
	volume = {60},
	issn = {0029-5515, 1741-4326},
	url = {https://iopscience.iop.org/article/10.1088/1741-4326/ab8a6a},
	doi = {10.1088/1741-4326/ab8a6a},
	number = {6},
	urldate = {2023-03-24},
	journal = {Nucl. Fusion},
	author = {Reimerdes, H. and Ambrosino, R. and Innocente, P. and Castaldo, A. and Chmielewski, P. and Di Gironimo, G. and Merriman, S. and Pericoli-Ridolfini, V. and Aho-Mantilla, L. and Albanese, R. and Bufferand, H. and Calabro, G. and Ciraolo, G. and Coster, D. and Fedorczak, N. and Ha, S. and Kembleton, R. and Lackner, K. and Loschiavo, V.P. and Lunt, T. and Marzullo, D. and Maurizio, R. and Militello, F. and Ramogida, G. and Subba, F. and Varoutis, S. and Zagórski, R. and Zohm, H.},
	month = jun,
	year = {2020},
	pages = {066030},
}

@article{kuang_divertor_2020,
	title = {Divertor heat flux challenge and mitigation in {SPARC}},
	volume = {86},
	issn = {0022-3778, 1469-7807},
	url = {https://www.cambridge.org/core/product/identifier/S0022377820001117/type/journal_article},
	doi = {10.1017/S0022377820001117},
	language = {en},
	number = {5},
	urldate = {2023-03-20},
	journal = {J. Plasma Phys.},
	author = {Kuang, A. Q. and Ballinger, S. and Brunner, D. and Canik, J. and Creely, A. J. and Gray, T. and Greenwald, M. and Hughes, J. W. and Irby, J. and LaBombard, B. and Lipschultz, B. and Lore, J. D. and Reinke, M. L. and Terry, J. L. and Umansky, M. and Whyte, D. G. and Wukitch, S. and {the SPARC Team}},
	month = oct,
	year = {2020},
	pages = {865860505},
}

@article{kuang_conceptual_2018,
	title = {Conceptual design study for heat exhaust management in the {ARC} fusion pilot plant},
	volume = {137},
	issn = {09203796},
	url = {https://linkinghub.elsevier.com/retrieve/pii/S0920379618306185},
	doi = {10.1016/j.fusengdes.2018.09.007},
	language = {en},
	urldate = {2023-03-14},
	journal = {Fusion Engineering and Design},
	author = {Kuang, A.Q. and Cao, N.M. and Creely, A.J. and Dennett, C.A. and Hecla, J. and LaBombard, B. and Tinguely, R.A. and Tolman, E.A. and Hoffman, H. and Major, M. and Ruiz Ruiz, J. and Brunner, D. and Grover, P. and Laughman, C. and Sorbom, B.N. and Whyte, D.G.},
	month = dec,
	year = {2018},
	pages = {221--242},
}

@article{umansky_attainment_2017,
	title = {Attainment of a stable, fully detached plasma state in innovative divertor configurations},
	volume = {24},
	issn = {1070-664X, 1089-7674},
	url = {http://aip.scitation.org/doi/10.1063/1.4979193},
	doi = {10.1063/1.4979193},
	language = {en},
	number = {5},
	urldate = {2023-03-14},
	journal = {Physics of Plasmas},
	author = {Umansky, M. V. and LaBombard, B. and Brunner, D. and Rensink, M. E. and Rognlien, T. D. and Terry, J. L. and Whyte, D. G.},
	month = may,
	year = {2017},
	pages = {056112},
}

@article{eich_scaling_2013,
	title = {Scaling of the tokamak near the scrape-off layer {H}-mode power width and implications for {ITER}},
	volume = {53},
	issn = {0029-5515, 1741-4326},
	url = {https://iopscience.iop.org/article/10.1088/0029-5515/53/9/093031},
	doi = {10.1088/0029-5515/53/9/093031},
	number = {9},
	urldate = {2023-03-12},
	journal = {Nucl. Fusion},
	author = {Eich, T. and Leonard, A.W. and Pitts, R.A. and Fundamenski, W. and Goldston, R.J. and Gray, T.K. and Herrmann, A. and Kirk, A. and Kallenbach, A. and Kardaun, O. and Kukushkin, A.S. and LaBombard, B. and Maingi, R. and Makowski, M.A. and Scarabosio, A. and Sieglin, B. and Terry, J. and Thornton, A. and {ASDEX Upgrade Team} and {JET EFDA Contributors}},
	month = sep,
	year = {2013},
	pages = {093031},
}

@article{kallenbach_impurity_2013,
	title = {Impurity seeding for tokamak power exhaust: from present devices via {ITER} to {DEMO}},
	volume = {55},
	issn = {0741-3335, 1361-6587},
	shorttitle = {Impurity seeding for tokamak power exhaust},
	url = {https://iopscience.iop.org/article/10.1088/0741-3335/55/12/124041},
	doi = {10.1088/0741-3335/55/12/124041},
	number = {12},
	urldate = {2023-02-10},
	journal = {Plasma Phys. Control. Fusion},
	author = {Kallenbach, A and Bernert, M and Dux, R and Casali, L and Eich, T and Giannone, L and Herrmann, A and McDermott, R and Mlynek, A and Müller, H W and Reimold, F and Schweinzer, J and Sertoli, M and Tardini, G and Treutterer, W and Viezzer, E and Wenninger, R and Wischmeier, M and {the ASDEX Upgrade Team}},
	month = dec,
	year = {2013},
	pages = {124041},
}

@article{reimerdes_initial_2021,
	title = {Initial {TCV} operation with a baffled divertor},
	volume = {61},
	issn = {0029-5515, 1741-4326},
	url = {https://iopscience.iop.org/article/10.1088/1741-4326/abd196},
	doi = {10.1088/1741-4326/abd196},
	number = {2},
	urldate = {2023-02-10},
	journal = {Nucl. Fusion},
	author = {Reimerdes, H. and Duval, B.P. and Elaian, H. and Fasoli, A. and Février, O. and Theiler, C. and Bagnato, F. and Baquero-Ruiz, M. and Blanchard, P. and Brida, D. and Colandrea, C. and De Oliveira, H. and Galassi, D. and Gorno, S. and Henderson, S. and Komm, M. and Linehan, B. and Martinelli, L. and Maurizio, R. and Moret, J.-M. and Perek, A. and Raj, H. and Sheikh, U. and Testa, D. and Toussaint, M. and Tsui, C.K. and Wensing, M. and TCV team, the and EUROfusion MST1 team, the},
	month = feb,
	year = {2021},
	pages = {024002},
}

@article{gorno_power_2023,
	title = {Power exhaust and core-divertor compatibility of the baffled snowflake divertor in {TCV}},
	volume = {65},
	issn = {0741-3335, 1361-6587},
	url = {https://iopscience.iop.org/article/10.1088/1361-6587/acad26},
	doi = {10.1088/1361-6587/acad26},
	number = {3},
	urldate = {2023-02-10},
	journal = {Plasma Phys. Control. Fusion},
	author = {Gorno, S and Colandrea, C and Février, O and Reimerdes, H and Theiler, C and Duval, B P and Lunt, T and Raj, H and Sheikh, U A and Simons, L and Thornton, A and Team, The TCV and MST1 Team, The EUROfusion},
	month = mar,
	year = {2023},
	pages = {035004},
}

@article{koenders_model-based_2023,
	title = {Model-based impurity emission front control using deuterium fueling and nitrogen seeding in {TCV}},
	volume = {63},
	issn = {0029-5515, 1741-4326},
	url = {https://iopscience.iop.org/article/10.1088/1741-4326/aca620},
	doi = {10.1088/1741-4326/aca620},
	number = {2},
	urldate = {2023-02-10},
	journal = {Nucl. Fusion},
	author = {Koenders, J.T.W. and Perek, A. and Kool, B. and Février, O. and Ravensbergen, T. and Galperti, C. and Duval, B.P. and Theiler, C. and van Berkel, M.},
	month = feb,
	year = {2023},
	pages = {026006},
}

@article{raj_improved_2022,
	title = {Improved heat and particle flux mitigation in high core confinement, baffled, alternative divertor configurations in the {TCV} tokamak},
	volume = {62},
	issn = {0029-5515, 1741-4326},
	url = {https://iopscience.iop.org/article/10.1088/1741-4326/ac94e5},
	doi = {10.1088/1741-4326/ac94e5},
	number = {12},
	urldate = {2023-02-10},
	journal = {Nucl. Fusion},
	author = {Raj, Harshita and Theiler, C. and Thornton, A. and Février, O. and Gorno, S. and Bagnato, F. and Blanchard, P. and Colandrea, C. and de Oliveira, H. and Duval, B.P. and Labit, B. and Perek, A. and Reimerdes, H. and Sheikh, U. and Vallar, M. and Vincent, B.},
	month = dec,
	year = {2022},
	pages = {126035},
}

@article{hofmann_creation_1994,
	title = {Creation and control of variably shaped plasmas in {TCV}},
	volume = {36},
	issn = {0741-3335, 1361-6587},
	url = {https://iopscience.iop.org/article/10.1088/0741-3335/36/12B/023},
	doi = {10.1088/0741-3335/36/12B/023},
	number = {12B},
	urldate = {2023-02-08},
	journal = {Plasma Phys. Control. Fusion},
	author = {Hofmann, F and Lister, J B and Anton, W and Barry, S and Behn, R and Bernel, S and Besson, G and Buhlmann, F and Chavan, R and Corboz, M and Dutch, M J and Duval, B P and Fasel, D and Favre, A and Franke, S and Heym, A and Hirt, A and Hollenstein, C and Isoz, P and Joye, B and Llobet, X and Magnin, J C and Marletaz, B and Marmillod, P and Martin, Y and Mayor, J M and Moret, J M and Nieswand, C and Paris, P J and Perez, A and Pietrzyk, Z A and Pitts, R A and Pochelon, A and Rage, R and Sauter, O and Tonetti, G and Tran, M Q and Troyon, F and Ward, D J and Weisen, H},
	month = dec,
	year = {1994},
	pages = {B277--B287},
}

@article{maurizio_divertor_2018,
	title = {Divertor power load studies for attached {L}-mode single-null plasmas in {TCV}},
	volume = {58},
	issn = {0029-5515, 1741-4326},
	url = {https://iopscience.iop.org/article/10.1088/1741-4326/aa986b},
	doi = {10.1088/1741-4326/aa986b},
	number = {1},
	urldate = {2023-01-16},
	journal = {Nucl. Fusion},
	author = {Maurizio, R. and Elmore, S. and Fedorczak, N. and Gallo, A. and Reimerdes, H. and Labit, B. and Theiler, C. and Tsui, C.K. and Vijvers, W.A.J. and {The TCV Team} and {The MST1 Team}},
	month = jan,
	year = {2018},
	pages = {016052},
}

@article{potzel_new_2014,
	title = {A new experimental classification of divertor detachment in {ASDEX} {Upgrade}},
	volume = {54},
	issn = {0029-5515, 1741-4326},
	url = {https://iopscience.iop.org/article/10.1088/0029-5515/54/1/013001},
	doi = {10.1088/0029-5515/54/1/013001},
	number = {1},
	urldate = {2023-01-11},
	journal = {Nucl. Fusion},
	author = {Potzel, S. and Wischmeier, M. and Bernert, M. and Dux, R. and Müller, H.W. and Scarabosio, A. and {the ASDEX Upgrade Team}},
	month = jan,
	year = {2014},
	pages = {013001},
}

@article{krasheninnikov_physics_2017,
	title = {Physics of ultimate detachment of a tokamak divertor plasma},
	volume = {83},
	issn = {0022-3778, 1469-7807},
	url = {https://www.cambridge.org/core/product/identifier/S0022377817000654/type/journal_article},
	doi = {10.1017/S0022377817000654},
	language = {en},
	number = {5},
	urldate = {2022-10-18},
	journal = {J. Plasma Phys.},
	author = {Krasheninnikov, S. I. and Kukushkin, A. S.},
	month = oct,
	year = {2017},
	pages = {155830501},
}

@article{goldston_new_2017,
	title = {A new scaling for divertor detachment},
	volume = {59},
	issn = {0741-3335, 1361-6587},
	url = {https://iopscience.iop.org/article/10.1088/1361-6587/aa5e6e},
	doi = {10.1088/1361-6587/aa5e6e},
	number = {5},
	urldate = {2022-10-03},
	journal = {Plasma Phys. Control. Fusion},
	author = {Goldston, R J and Reinke, M L and Schwartz, J A},
	month = may,
	year = {2017},
	pages = {055015},
}

@article{reimerdes_tcv_2017,
	title = {{TCV} experiments towards the development of a plasma exhaust solution},
	volume = {57},
	issn = {0029-5515, 1741-4326},
	url = {https://iopscience.iop.org/article/10.1088/1741-4326/aa82c2},
	doi = {10.1088/1741-4326/aa82c2},
	number = {12},
	urldate = {2022-09-02},
	journal = {Nucl. Fusion},
	author = {Reimerdes, H. and Duval, B.P. and Harrison, J.R. and Labit, B. and Lipschultz, B. and Lunt, T. and Theiler, C. and Tsui, C.K. and Verhaegh, K. and Vijvers, W.A.J. and Boedo, J.A. and Calabro, G. and Crisanti, F. and Innocente, P. and Maurizio, R. and Pericoli, V. and Sheikh, U. and Spolare, M. and Vianello, N. and {the TCV team} and {the EUROfusion MST1 team}},
	month = dec,
	year = {2017},
	pages = {126007},
}

@article{pitts_divertor_2001,
	series = {14th {Int}. {Conf}. on {Plasma}-{Surface} {Interactions} in {Controlled} {Fusion} {D} evices},
	title = {Divertor geometry effects on detachment in {TCV}},
	volume = {290-293},
	issn = {0022-3115},
	url = {https://www.sciencedirect.com/science/article/pii/S002231150000461X},
	doi = {10.1016/S0022-3115(00)00461-X},
	language = {en},
	urldate = {2022-08-31},
	journal = {Journal of Nuclear Materials},
	author = {Pitts, R. A. and Duval, B. P. and Loarte, A. and Moret, J. -M. and Boedo, J. A. and Coster, D. and Furno, I. and Horacek, J. and Kukushkin, A. S. and Reiter, D. and Rommers, J.},
	month = mar,
	year = {2001},
	pages = {940--946},
}

@article{lipschultz_sensitivity_2016,
	title = {Sensitivity of detachment extent to magnetic configuration and external parameters},
	volume = {56},
	issn = {0029-5515, 1741-4326},
	url = {https://iopscience.iop.org/article/10.1088/0029-5515/56/5/056007},
	doi = {10.1088/0029-5515/56/5/056007},
	number = {5},
	urldate = {2022-08-31},
	journal = {Nucl. Fusion},
	author = {Lipschultz, Bruce and Parra, Felix I. and Hutchinson, Ian H.},
	month = may,
	year = {2016},
	pages = {056007},
}

@article{kotschenreuther_heat_2007,
	title = {On heat loading, novel divertors, and fusion reactors},
	volume = {14},
	issn = {1070-664X},
	url = {https://aip.scitation.org/doi/10.1063/1.2739422},
	doi = {10.1063/1.2739422},
	number = {7},
	urldate = {2022-08-30},
	journal = {Physics of Plasmas},
	author = {Kotschenreuther, M. and Valanju, P. M. and Mahajan, S. M. and Wiley, J. C.},
	month = jul,
	year = {2007},
	pages = {072502},
}

@article{labombard_adx_2015,
	title = {{ADX}: a high field, high power density, advanced divertor and {RF} tokamak},
	volume = {55},
	issn = {0029-5515, 1741-4326},
	shorttitle = {{ADX}},
	url = {https://iopscience.iop.org/article/10.1088/0029-5515/55/5/053020},
	doi = {10.1088/0029-5515/55/5/053020},
	number = {5},
	urldate = {2022-08-30},
	journal = {Nucl. Fusion},
	author = {LaBombard, B. and Marmar, E. and Irby, J. and Terry, J.L. and Vieira, R. and Wallace, G. and Whyte, D.G. and Wolfe, S. and Wukitch, S. and Baek, S. and Beck, W. and Bonoli, P. and Brunner, D. and Doody, J. and Ellis, R. and Ernst, D. and Fiore, C. and Freidberg, J.P. and Golfinopoulos, T. and Granetz, R. and Greenwald, M. and Hartwig, Z.S. and Hubbard, A. and Hughes, J.W. and Hutchinson, I.H. and Kessel, C. and Kotschenreuther, M. and Leccacorvi, R. and Lin, Y. and Lipschultz, B. and Mahajan, S. and Minervini, J. and Mumgaard, R. and Nygren, R. and Parker, R. and Poli, F. and Porkolab, M. and Reinke, M.L. and Rice, J. and Rognlien, T. and Rowan, W. and Shiraiwa, S. and Terry, D. and Theiler, C. and Titus, P. and Umansky, M. and Valanju, P. and Walk, J. and White, A. and Wilson, J.R. and Wright, G. and Zweben, S.J.},
	month = may,
	year = {2015},
	pages = {053020},
}

@article{eich_inter-elm_2011,
	title = {Inter-{ELM} {Power} {Decay} {Length} for {JET} and {ASDEX} {Upgrade}: {Measurement} and {Comparison} with {Heuristic} {Drift}-{Based} {Model}},
	volume = {107},
	issn = {0031-9007, 1079-7114},
	shorttitle = {Inter-{ELM} {Power} {Decay} {Length} for {JET} and {ASDEX} {Upgrade}},
	url = {https://link.aps.org/doi/10.1103/PhysRevLett.107.215001},
	doi = {10.1103/PhysRevLett.107.215001},
	language = {en},
	number = {21},
	urldate = {2022-08-30},
	journal = {Phys. Rev. Lett.},
	author = {Eich, T. and Sieglin, B. and Scarabosio, A. and Fundamenski, W. and Goldston, R. J. and Herrmann, A. and {ASDEX Upgrade Team}},
	month = nov,
	year = {2011},
	pages = {215001},
}

@article{fevrier_divertor_2021,
	title = {Divertor closure effects on the {TCV} boundary plasma},
	volume = {27},
	issn = {2352-1791},
	url = {https://www.sciencedirect.com/science/article/pii/S2352179121000600},
	doi = {10.1016/j.nme.2021.100977},
	language = {en},
	urldate = {2022-08-30},
	journal = {Nuclear Materials and Energy},
	author = {Février, O. and Reimerdes, H. and Theiler, C. and Brida, D. and Colandrea, C. and De Oliveira, H. and Duval, B. P. and Galassi, D. and Gorno, S. and Henderson, S. and Komm, M. and Labit, B. and Linehan, B. and Martinelli, L. and Perek, A. and Raj, H. and Sheikh, U. and Tsui, C. K. and Wensing, M.},
	month = jun,
	year = {2021},
	pages = {100977},
}

@article{fevrier_detachment_2021,
	title = {Detachment in conventional and advanced double-null plasmas in {TCV}},
	volume = {61},
	issn = {0029-5515, 1741-4326},
	url = {https://iopscience.iop.org/article/10.1088/1741-4326/ac27c6},
	doi = {10.1088/1741-4326/ac27c6},
	number = {11},
	urldate = {2022-08-30},
	journal = {Nucl. Fusion},
	author = {Février, O. and Theiler, C. and Coda, S. and Colandrea, C. and de Oliveira, H. and Duval, B.P. and Gorno, S. and Labit, B. and Linehan, B. and Maurizio, R. and Perek, A. and Reimerdes, H. and Wüthrich, C. and {the TCV Team}},
	month = nov,
	year = {2021},
	pages = {116064},
}

@article{theiler_results_2017,
	title = {Results from recent detachment experiments in alternative divertor configurations on {TCV}},
	volume = {57},
	issn = {0029-5515, 1741-4326},
	url = {https://iopscience.iop.org/article/10.1088/1741-4326/aa5fb7},
	doi = {10.1088/1741-4326/aa5fb7},
	number = {7},
	urldate = {2022-08-30},
	journal = {Nucl. Fusion},
	author = {Theiler, C. and Lipschultz, B. and Harrison, J. and Labit, B. and Reimerdes, H. and Tsui, C. and Vijvers, W.A.J. and Boedo, J. A. and Duval, B.P. and Elmore, S. and Innocente, P. and Kruezi, U. and Lunt, T. and Maurizio, R. and Nespoli, F. and Sheikh, U. and Thornton, A.J. and van Limpt, S.H.M. and Verhaegh, K. and Vianello, N.},
	month = jul,
	year = {2017},
	pages = {072008},
}

@article{maurizio_h-mode_2021,
	title = {H-mode scrape-off layer power width in the {TCV} tokamak},
	volume = {61},
	issn = {0029-5515, 1741-4326},
	url = {https://iopscience.iop.org/article/10.1088/1741-4326/abd147},
	doi = {10.1088/1741-4326/abd147},
	number = {2},
	urldate = {2022-08-30},
	journal = {Nucl. Fusion},
	author = {Maurizio, R. and Duval, B.P. and Labit, B. and Reimerdes, H. and Faitsch, M. and Komm, M. and Sheikh, U. and Theiler, C. and {the TCV team}},
	month = feb,
	year = {2021},
	pages = {024003},
}

@article{kotschenreuther_magnetic_2013,
	title = {Magnetic geometry and physics of advanced divertors: {The} {X}-divertor and the snowflake},
	volume = {20},
	issn = {1070-664X, 1089-7674},
	shorttitle = {Magnetic geometry and physics of advanced divertors},
	url = {https://pubs.aip.org/pop/article/20/10/102507/318117/Magnetic-geometry-and-physics-of-advanced},
	doi = {10.1063/1.4824735},
	language = {en},
	number = {10},
	urldate = {2023-08-29},
	journal = {Physics of Plasmas},
	author = {Kotschenreuther, Mike and Valanju, Prashant and Covele, Brent and Mahajan, Swadesh},
	month = oct,
	year = {2013},
	pages = {102507},
}

@article{sun_performance_2023,
	title = {Performance assessment of a tightly baffled, long-legged divertor configuration in {TCV} with {SOLPS}-{ITER}},
	volume = {63},
	issn = {0029-5515, 1741-4326},
	url = {https://iopscience.iop.org/article/10.1088/1741-4326/ace45f},
	doi = {10.1088/1741-4326/ace45f},
	number = {9},
	urldate = {2023-08-21},
	journal = {Nucl. Fusion},
	author = {Sun, G. and Reimerdes, H. and Theiler, C. and Duval, B.P. and Carpita, M. and Colandrea, C. and Février, O.},
	month = sep,
	year = {2023},
	pages = {096011},
}

@article{wang_first_2020,
	title = {First {Evidence} of {Local} {E} × {B} {Drift} in the {Divertor} {Influencing} the {Structure} and {Stability} of {Confined} {Plasma} near the {Edge} of {Fusion} {Devices}},
	volume = {124},
	issn = {0031-9007, 1079-7114},
	url = {https://link.aps.org/doi/10.1103/PhysRevLett.124.195002},
	doi = {10.1103/PhysRevLett.124.195002},
	language = {en},
	number = {19},
	urldate = {2023-08-16},
	journal = {Phys. Rev. Lett.},
	author = {Wang, H. Q. and Guo, H. Y. and Xu, G. S. and Leonard, A. W. and Wu, X. Q. and Groth, M. and Jaervinen, A. E. and Watkins, J. G. and Osborne, T. H. and Thomas, D. M. and Eldon, D. and Stangeby, P. C. and Turco, F. and Xu, J. C. and Wang, L. and Wang, Y. F. and Liu, J. B.},
	month = may,
	year = {2020},
	pages = {195002},
}

@article{leonard_plasma_2018,
	title = {Plasma detachment in divertor tokamaks},
	volume = {60},
	issn = {0741-3335, 1361-6587},
	url = {https://iopscience.iop.org/article/10.1088/1361-6587/aaa7a9},
	doi = {10.1088/1361-6587/aaa7a9},
	number = {4},
	urldate = {2023-08-13},
	journal = {Plasma Phys. Control. Fusion},
	author = {Leonard, A W},
	month = apr,
	year = {2018},
	pages = {044001},
}

@article{pitts_experimental_1999,
	title = {Experimental investigation of the effects of neon injection in {TCV}},
	volume = {266-269},
	issn = {00223115},
	url = {https://linkinghub.elsevier.com/retrieve/pii/S0022311598005996},
	doi = {10.1016/S0022-3115(98)00599-6},
	language = {en},
	urldate = {2023-08-11},
	journal = {Journal of Nuclear Materials},
	author = {Pitts, R.A. and Refke, A. and Duval, B.P. and Furno, I. and Joye, B. and Lister, J.B. and Martin, Y. and Moret, J.-M. and Rommers, J. and Weisen, H.},
	month = mar,
	year = {1999},
	pages = {648--653},
}

@article{ryutov_snowflake_2015,
	title = {The snowflake divertor},
	volume = {22},
	issn = {1070-664X, 1089-7674},
	url = {https://pubs.aip.org/aip/pop/article/108992},
	doi = {10.1063/1.4935115},
	language = {en},
	number = {11},
	urldate = {2023-06-14},
	journal = {Phys. Plasmas},
	author = {Ryutov, D. D. and Soukhanovskii, V. A.},
	month = nov,
	year = {2015},
	pages = {110901},
}

@article{petrie_effect_2013,
	title = {Effect of changes in separatrix magnetic geometry on divertor behaviour in {DIII}-{D}},
	volume = {53},
	issn = {0029-5515, 1741-4326},
	url = {https://iopscience.iop.org/article/10.1088/0029-5515/53/11/113024},
	doi = {10.1088/0029-5515/53/11/113024},
	number = {11},
	urldate = {2023-04-19},
	journal = {Nucl. Fusion},
	author = {Petrie, T.W. and Canik, J.M. and Lasnier, C.J. and Leonard, A.W. and Mahdavi, M.A. and Watkins, J.G. and Fenstermacher, M.E. and Ferron, J.R. and Groebner, R.J. and Hill, D.N. and Hyatt, A.W. and Holcomb, C.T. and Luce, T.C. and Makowski, M. and Moyer, R.A. and Osborne, T.E. and Stangeby, P.C.},
	month = nov,
	year = {2013},
	pages = {113024},
}

@article{reinke_heat_2017,
	title = {Heat flux mitigation by impurity seeding in high-field tokamaks},
	volume = {57},
	issn = {0029-5515, 1741-4326},
	url = {https://iopscience.iop.org/article/10.1088/1741-4326/aa5145},
	doi = {10.1088/1741-4326/aa5145},
	number = {3},
	urldate = {2023-04-18},
	journal = {Nucl. Fusion},
	author = {Reinke, M.L.},
	month = mar,
	year = {2017},
	pages = {034004},
}

@article{moulton_comparison_2021,
	title = {Comparison between {SOLPS}-4.3 and the {Lengyel} {Model} for {ITER} baseline neon-seeded plasmas},
	volume = {61},
	issn = {0029-5515, 1741-4326},
	url = {https://iopscience.iop.org/article/10.1088/1741-4326/abe4b2},
	doi = {10.1088/1741-4326/abe4b2},
	number = {4},
	urldate = {2023-04-18},
	journal = {Nucl. Fusion},
	author = {Moulton, D. and Stangeby, P.C. and Bonnin, X. and Pitts, R.A.},
	month = apr,
	year = {2021},
	pages = {046029},
}

@article{fasoli_tcv_2020,
	title = {{TCV} heating and divertor upgrades},
	volume = {60},
	issn = {0029-5515, 1741-4326},
	url = {https://iopscience.iop.org/article/10.1088/1741-4326/ab4c56},
	doi = {10.1088/1741-4326/ab4c56},
	number = {1},
	urldate = {2023-03-27},
	journal = {Nucl. Fusion},
	author = {Fasoli, A. and Reimerdes, H. and Alberti, S. and Baquero-Ruiz, M. and Duval, B.P. and Havlikova, E. and Karpushov, A. and Moret, J.-M. and Toussaint, M. and Elaian, H. and Silva, M. and Theiler, C. and Vaccaro, D. and {the TCV team}},
	month = jan,
	year = {2020},
	pages = {016019},
}

@article{wigram_performance_2019,
	title = {Performance assessment of long-legged tightly-baffled divertor geometries in the {ARC} reactor concept},
	volume = {59},
	issn = {0029-5515, 1741-4326},
	url = {https://iopscience.iop.org/article/10.1088/1741-4326/ab394f},
	doi = {10.1088/1741-4326/ab394f},
	number = {10},
	urldate = {2023-03-27},
	journal = {Nucl. Fusion},
	author = {Wigram, M.R.K. and LaBombard, B. and Umansky, M.V. and Kuang, A.Q. and Golfinopoulos, T. and Terry, J.L. and Brunner, D. and Rensink, M.E. and Ridgers, C.P. and Whyte, D.G.},
	month = oct,
	year = {2019},
	pages = {106052},
}

@article{takase_guidance_2001,
	title = {Guidance of {Divertor} {Channel} by {Cusp}-{Like} {Magnetic} {Field} for {Tokamak} {Devices}},
	volume = {70},
	issn = {0031-9015, 1347-4073},
	url = {http://journals.jps.jp/doi/10.1143/JPSJ.70.609},
	doi = {10.1143/JPSJ.70.609},
	language = {en},
	number = {3},
	urldate = {2023-03-27},
	journal = {J. Phys. Soc. Jpn.},
	author = {Takase, Haruhiko},
	month = mar,
	year = {2001},
	pages = {609--612},
}

@article{valanju_super_2010,
	title = {Super {X} divertors for solving heat and neutron flux problems of fusion devices},
	volume = {85},
	issn = {09203796},
	url = {https://linkinghub.elsevier.com/retrieve/pii/S0920379609002439},
	doi = {10.1016/j.fusengdes.2009.06.001},
	language = {en},
	number = {1},
	urldate = {2023-03-27},
	journal = {Fusion Engineering and Design},
	author = {Valanju, P.M. and Kotschenreuther, M. and Mahajan, S.M.},
	month = jan,
	year = {2010},
	pages = {46--52},
}

@article{loarte_plasma_1998,
	title = {Plasma detachment in {JET} {Mark} {I} divertor experiments},
	volume = {38},
	issn = {0029-5515},
	url = {https://iopscience.iop.org/article/10.1088/0029-5515/38/3/303},
	doi = {10.1088/0029-5515/38/3/303},
	number = {3},
	urldate = {2023-09-27},
	journal = {Nucl. Fusion},
	author = {Loarte, A and Monk, R.D and Martín-Solís, J.R and Campbell, D.J and Chankin, A.V and Clement, S and Davies, S.J and Ehrenberg, J and Erents, S.K and Guo, H.Y and Harbour, P.J and Horton, L.D and Ingesson, L.C and Jäckel, H and Lingertat, J and Lowry, C.G and Maggi, C.F and Matthews, G.F and McCormick, K and O'Brien, D.P and Reichle, R and Saibene, G and Smith, R.J and Stamp, M.F and Stork, D and Vlases, G.C},
	month = mar,
	year = {1998},
	pages = {331--371},
}

\end{document}